\documentclass{aa}  
\relax
\usepackage{graphicx}
\usepackage{txfonts}
\usepackage{pdflscape}
\usepackage{color}

\usepackage{amstext}
\usepackage{amsmath}
\usepackage{amssymb}

\begin{document} 

\title{A technique to select the most obscured galaxy nuclei}

   \author{I. Garc\'ia-Bernete\inst{1}\fnmsep\thanks{E-mail: igbernete@gmail.com}, D. Rigopoulou\inst{1}, S. Aalto\inst{2}, H.W.W. Spoon\inst{3}, A. Hern\'an-Caballero\inst{4}, A. Efstathiou\inst{5}, P.F. Roche\inst{1} and S. K\"onig\inst{2}\\}

   \institute{$^1$Department of Physics, University of Oxford, Keble Road, Oxford OX1 3RH, UK \\
   $^2$Department of Space, Earth and Environment, Osala Space Observatory, Chalmers University of Technology, SE-439 92 Onsala, Sweden\\ 
   $^3$Cornell Center for Astrophysics and Planetary Science (CCAPS), Department of Astronomy, Cornell University, Ithaca, NY 14853\\
   $^4$Centro de Estudios de F\'isica del Cosmos de Arag\'on, Plaza San Juan, 1, E-44001 Teruel, Spain\\
   $^5$School of Sciences, European University Cyprus, Diogenes street, Engomi, 1516 Nicosia, Cyprus\\}

\titlerunning{A technique to select deeply buried nuclei}
\authorrunning{Garc\'ia-Bernete et al.}

   \date{}

  \abstract
   {Compact obscured nuclei (CONs) are mainly found in local luminous and ultraluminous infrared galaxies (U/LIRGs). In the local Universe, these sources are generally selected through the detection of the HCN--vib (3-2) emission line at submillimetre wavelengths. In this work, we present a diagnostic method to select deeply buried nuclei based on mid-infrared (mid-IR) polycyclic aromatic hydrocarbons (PAHs) and mid-IR continuum ratios. Using Spitzer InfraRed Spectrograph spectra of a representative sample of local ULIRGs (z$<$0.27), we examine their PAH and underlying continuum emission ratios. For deeply embedded sources, we find that the 9.7\,$\mu$m silicate absorption band has a particularly pronounced effect on the 11.3\,$\mu$m PAH feature. The low flux level in the nuclear silicate absorption band enhances the 11.3\,$\mu$m PAH feature contrast (high PAH equivalent width) compared to that of the other PAH features. The technique has been extended to include the use of the underlying 11.3/12.7 and 11.3/6.2~$\mu$m continuum ratios. However, the latter are affected both by the extinction coming from the host galaxy as well as the nuclear region, whereas the foreground (host-galaxy) extinction is cancelled out when using the PAH EW ratios. We apply our method to local U/LIRGs from the HERUS and GOALS samples and classify as CON candidates 14 ULIRGs and 10 LIRGs, corresponding to 30\% of ULIRGs and 7\% of LIRGs from these samples. We find that the observed continuum ratios of CON-dominated sources can be explained by assuming torus models with a tapered disk geometry and a smooth dust distribution. This suggests that the nuclear dusty structure of deeply obscured galaxy nuclei has an extremely high dust coverage. Finally, we demonstrate that the use of mid-IR color–color diagrams is an effective way to select CON-dominated sources at different redshifts. In particular, the combination of filters of the James Webb Space Telescope/Mid-Infrared Instrument will enable the selection of CONs out to z$\sim$1.5. This will allow extending the selection of CONs to high redshifts where U/LIRGs are more numerous.}

      
   \maketitle


\section{Introduction}
Recent submillimetre (sub-mm) observations have revealed the presence of compact (few tens of pc) and highly obscured nuclei (N$_{\rm H}>$10$^{25}$~cm$^{-2}$) in local luminous (10$^{11}<$L$_{\rm IR}<$
10$^{12}$ L$_{\odot}$, LIRGs) and ultraluminous ( $L_{\rm IR}>$
10$^{12}$ L$_{\odot}$, ULIRGs) infrared galaxies (e.g. \citealt{Aalto15} and references therein). 
These compact obscured nuclei (CONs) are considered to be an important phase of galaxy evolution (e.g. \citealt{Aalto15}), but because of the high obscuration, the nature of the nuclear embedded source is unclear: it may be an accreting supermassive black hole (SMBH) in an active galactic nucleus (AGN), a nuclear starburst, or a combination of both (e.g. \citealt{Veilleux09}). Regardless of the exact nature of the underlying source, studying the properties of CONs enables us to better understand growth processes in galaxies as well as to investigate the relation between black hole mass and spheroid growth (e.g. \citealt{Kormendy13}).
 
The extremely high extinction associated with the large column densities of gas and dust present in the inner regions of these sources impede their detection at many wavelengths (e.g. optical/X-ray). Earlier studies have routinely used the 9.7\,$\mu m$ silicate absorption band (e.g. \citealt{Imanishi07,Levenson07,Sajina07,Sajina09,Spoon07,Sirocky08,Georgantopoulos11}) and mid-IR colors (e.g. \citealt{Stern12,Mateos13}) for selecting heavily obscured sources. The high column densities of dust, however, produce large optical depths in the 9.7~$\mu$m silicate absorption band (e.g. \citealt{Ossenkopf92}). The dusty core absorbs a significant fraction of the intrinsic radiation and reprocesses it to emerge at longer wavelengths, in the IR. In particular, it has been found that IR emission of deeply buried U/LIRGs can be modelled by assuming dust surrounding the nuclear source with an additional contribution from a starburst and a spheroidal host (see e.g. \citealt{Herrero-Illana17,Marshall18,Efstathiou21} and references therein).

More recent works found that the HCN (3-2) line in its vibrationally excited $\nu_2$=1 state (HCN--vib) is an excellent tracer of CONs (e.g. \citealt{Sakamoto10,Imanishi13,Aalto15,Aalto19,Falstad21}). In particular,  \citet{Falstad21} suggested that a high HCN--vib surface brightness ($\Sigma_{\rm HCN-vib}>$1 L$_{\odot}$~pc$^{-2}$ and/or L$_{\rm HCN-vib}$/L$_{\rm IR}>$10$^{-8}$) indicates that the core may be optically thick up to sub-mm wavelengths (see also e.g. \citealt{Sakamoto13,Sakamoto17,Scoville17,Aalto19}). Using the HCN--vib molecular gas transition, previous studies have revealed that 20--40\% of the U/LIRGs in the local Universe harbour previously undetected CONs (e.g. \citealt{Sakamoto10,Imanishi13,Aalto15,Aalto19,Falstad21}). In particular, \citet{Falstad21} reported that CONs are more frequently found in ULIRGs ($\sim$38\%) than in LIRGs ($\sim$21\%) by using a volume-limited sample of 46 far-IR luminous galaxies (L$_{\rm FIR (40-400~\mu m)}>$10$^{10}$ L$_{\odot}$). However, more studies using larger samples are needed to confirm or otherwise whether this is a commonly occurring feature in U/LIRGs, its dependence on luminosity and$/$or other properties of the host galaxy.

Polycyclic Aromatic Hydrocarbons (PAH) features are ubiquitous in a variety of astrophysical objects and environments (see e.g. \citealt{Li20} for a review). PAH molecules absorb ultraviolet (UV)$/$optical photons resulting in their excitation. The excited molecules emit mid-IR photons via vibrational relaxation (e.g. \citealt{Leger84,Allamandola89}). PAH molecules are likely responsible for the family of strong infrared emission bands (3.3, 6.2, 7.7, 8.6, 11.3 and 12.7~$\mu$m; e.g. \citealt{Tielens08}) observed in galactic and extragalactic sources. PAH emission features are very prominent in the mid-IR spectra of U/LIRGs (e.g. \citealt{Lutz98,Rigopoulou99,Caballero11}). These molecules absorb a significant fraction of UV/optical photons (mainly) from young stars (e.g. \citealt{Peeters04}) and, therefore, are excellent tracers of the star formation activity in galaxies near and far (e.g. \citealt{Rigopoulou99,Peeters04,Brandl06,Smith07a,Farrah08,Pope08,Caballero9,Shipley16}).

PAH emission has been also detected in the IR spectra of CONs (e.g. \citealt{Falstad21}) by using the relatively large aperture afforded by Spitzer ($\sim$4\arcsec). However, for the archetypal CON  NGC\,4418, the 11.3~$\mu$m feature was not detected in its subarcsecond angular resolution ground-based mid-IR spectrum \citep{Roche15}. These authors also found a deeper silicate absorption band (i.e. lower flux level of the nuclear silicate absorption band minimum) compared to that measured in the total Spitzer/IRS spectrum. This suggests that in deeply embedded sources, the bulk of the observed PAH emission originates primarily in their circumnuclear regions and, therefore, is likely to experience substantially lower levels of extinction compared to the compact core. 

In this work, we use the Infrared Database of Extragalactic Observables from Spitzer (IDEOS database; \citealt{Hernan-Caballero16,Hernan-Caballero20,Spoon21}) and archival Spitzer IR spectral data of a representative sample of local U/LIRGs to investigate the impact 
of the nuclear 9.7\,$\mu$m silicate absorption on their galaxy integrated PAH emission. 
The paper is organized as follows: In Sections \ref{sample_selection} and \ref{observations} we describe the sample selection and observations used in this study, respectively. The main results of the PAH emission are presented in Section  \ref{pahdiagram}. A new diagnostic diagram for selecting deeply buried nuclei is discussed in Section \ref{Diagnostic_diagrams}. In Section \ref{filter}, we discuss the ability of James Webb Space Telescope (JWST) mid-infrared instrument (MIRI) imager to isolate deeply obscured nuclei. Finally, in Section \ref{conclusions}, we summarize the main conclusions of this work.

\section{The Samples}
\label{sample_selection}
We select a representative sample of local ULIRGs starting from the HERschel Ultra Luminous Infrared Galaxy Survey (HERUS; \citealt{Farrah13}), which was observed with the Herschel Space Observatory \citep{Pilbratt10} in spectroscopy and photometry mode. Although it is not a strictly flux-limited, we chose the HERUS sample since it includes nearly all ULIRGs at z$<$0.27 with 60~$\mu$m fluxes between 1.7 and 6 Jy. The sample consists of the 42 most luminous ULIRGs in the local Universe, which were also observed by the Spitzer/Infrared Spectrograph (IRS; \citealt{Houck04}). We refer the reader to \citet{Farrah13} and \citet{Pearson16} for further details on the sample.

Although our main focus is on ULIRGs, to sample a wide range of infrared luminosities, we also included LIRGs. To do so, we use the LIRGs sample from the Great Observatories All-sky LIRG Survey (GOALS; \citealt{Armus09}). The GOALS sample consist of a complete subset of the Infrared Astronomical Satellite (IRAS) Revised Bright Galaxy Sample (RBGS; \citealt{Sanders03}). The GOALS sample consists of 179 LIRGs and 22 ULIRGs, which cover different interaction stages and nuclear spectral types. Note that 18 of the 22 ULIRGs in the GOALS sample are in common with the HERUS sample. Thus, the combination of the HERUS and GOALS samples leaves a final sample of 46 ULIRGs and 179 LIRGs.

Finally, for the present study, we also use a sample of far-infrared luminous galaxies that include CONs (CON-quest sample; \citealt{Falstad21}). Although there is a significant overlap between the U/LIRGs present in the CON-quest sample and the samples previously described (see Table \ref{taboverlap}), we use the CON-quest sample for the sole purpose of testing the effectiveness of the technique for selecting CON sources. Note that \citet{Falstad21} classified as CON sources only a fraction of the CON-quest sample ($\sim$38\% of the ULIRGs and $\sim$21\% of the LIRGs), based on their high HCN–vib surface brightness (see \citealt{Falstad21} for further details).

\begin{table}
\centering
\begin{tabular}{lcc}
\hline
Type of source & HERUS   &	GOALS \\
 (1)&(2)&(3)\\	
\hline
ULIRGs & 7/8 & 8/8\\
LIRGs & 0/19 & 18/19\\
sub-LIRGs & 0/19& 0/19\\
\hline
\end{tabular}						 
\caption{Sources in common between the HERUS and GOALS samples and the CONquest sample. Note that the denominators correspond to ULIRGs, LIRGs and sub-LIRGs that are part of the CONquest sample.}
\label{taboverlap}
\end{table}

\begin{figure*}[ht!]
\centering
\includegraphics[width=19.6cm]{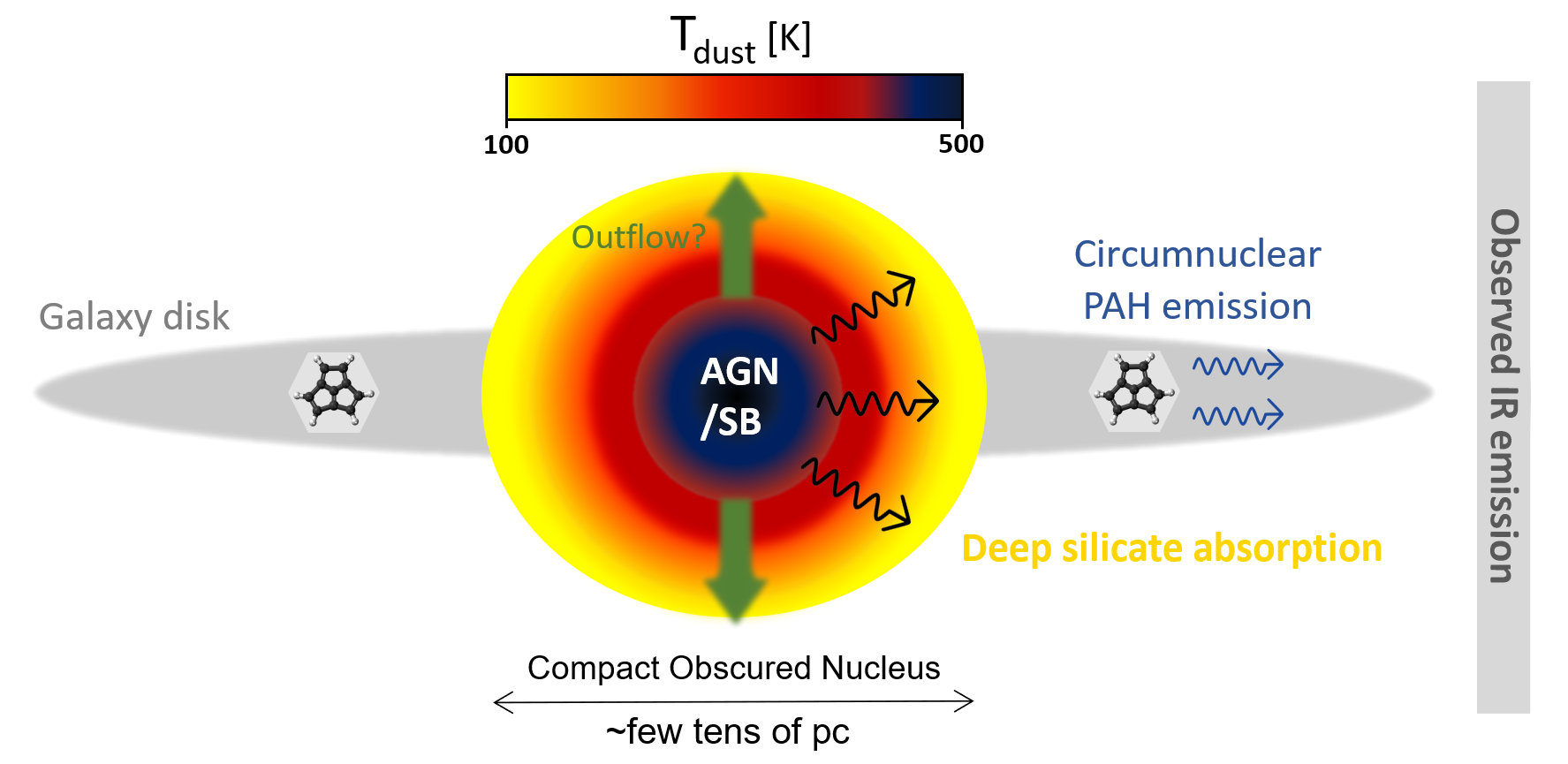}
\caption{A schematic view of a galaxy containing a CON (not to scale). The cartoon depicts a compact obscured nucleus, which its submillimetre HCN--vib emission is emerging from inside the inner few tens of pc \citep{Aalto15}, surrounded by a star-formation galaxy disk where PAH molecules mainly reside. The total IR emission of a CON galaxy includes contributions from two distinct components: one coming from a heavily obscured nucleus (with a deep 9.7~$\mu$m silicate absorption) and the other from the host galaxy disk (cicumnuclear PAH emission). The colour scheme used represents the dust temperature gradient of the CON ranging from black (high values) to yellow (lower values of the dust temperature). The green vertical arrows represent the existence of possible outflows whose presence has been confirmed in some CON sources (e.g. \citealt{Falstad19} and references therein) suggesting that the geometry of the compact obscured nucleus may not be perfectly spherically symmetric. The grey shaded rectangle on the right side of the plot (labelled as ``observed IR emission'') represents the large aperture of Spitzer/IRS which contains a significant part of the total IR flux of the galaxy. Note that the fraction of the galaxy included in the Spitzer/IRS aperture depend on the size of the slit with respect to the size of the galaxy and its distance (see Appendix \ref{fraction_irs}). See also Fig. \ref{Synth_con} for further details on the total IR emission of galaxies containing CONs.} 
\label{cartoon}
\end{figure*}

\section{Literature data}
\label{observations}

The samples described above
were cross-correlated with the Spitzer Heritage Archive (SHA)\footnote{https://sha.ipac.caltech.edu/applications/Spitzer/SHA/} in order to select only those with Spitzer IRS staring mode observations. In particular, we use the low (R$\sim$60--120) resolution IRS modules: Short-Low (SL1; 7.4--14.5~$\mu$m, SL2; 5.2--7.7~$\mu$m) and Long-Low (LL1; 19.5--38.0~$\mu$m, LL2; 14.0--21.3~$\mu$m).

For those galaxies, we retrieve 6.2, 11.3 and 12.7~$\mu$m PAH and continuum measurements, and 9.7~$\mu$m silicate strengths from the IDEOS database (\citealt{Spoon21}). The IDEOS spectra were drawn from the Cornell Atlas of Spitzer/IRS Source (CASSIS, version LR7; \citealt{Lebouteiller11}). The IDEOS database includes all the spectra from CASSIS for galaxies beyond the Local Group. The spectra were reduced with the CASSIS software, using optimal extraction and tapered column extraction depending on the source spatial extent (see \citealt{Lebouteiller11}). Note that to stitch together the different Spitzer/IRS modules it is needed to apply a small scaling factor. The SL spectrum is scaled to get a smooth stitch between SL and LL modules (see \citealt{Hernan-Caballero16} for a full description of the automated stitching routine used in IDEOS).

A full description of the methods used to obtain the IDEOS measurements are presented in \citet{Hernan-Caballero16,Hernan-Caballero20} and \citet{Spoon21}. In summary, PAH fluxes and equivalent widths (EWs) are measured in different spectral ranges by using a combination of polynomial continua, Pearson type-IV distribution profiles for the 6.2, 11.3 and 12.7\,$\mu$m PAH features and Gaussian profiles for the weaker PAH bands, fine-structure and H$_{2}$ emission lines. The models used also consider attenuation due to the presence of water ices in the 6.2\,$\mu$m PAH band region. Note that 
the model described above is not used for measuring 7.7 and 8.6\,$\mu$m PAHs because of the difficulty in determining a local continuum for these features. The silicate strength is computed as S$_{\rm Sil}=$ln(f$_{\rm peak}/$f$_{\rm cont}$), where f$_{\rm peak}$ corresponds to the peak of the emission or absorption silicate feature and f$_{\rm cont}$ is the flux of the underlying continuum measured at the wavelength of the peak. For the underlying continuum, IDEOS uses either a spline or power-law interpolation between anchor points on both sides of the silicate feature (see \citealt{Spoon21} for details).

\section{The mid-infrared emission of highly obscured dusty galaxies}
\label{pahdiagram}
Regardless of the exact nature of the hidden power source, \citet{Sirocky08} found that when the obscuration is sufficiently high, the radiation emitted by the central source (massive star, AGN and/or starburst nucleus) has a negligible effect on the spectral shape of the emerging IR emission from the surrounding dusty shell. Although PAH emission has been detected in the nuclear region of AGN using subarcsecond resolution data (e.g. \citealt{Hoenig10,Herrero14,Herrero16,Esquej14,Jensen17}), the nuclear PAH emission can be relatively weak compare with that of the circumnuclear regions due to the potential destruction of PAH molecules by the hard radiation field of the nuclear source (e.g. \citealt{Roche91,Voit92,Siebenmorgen04}). However, shielding of PAH molecules by molecular hydrogen (H$_{2}$) is also possible in these regions (e.g. \citealt{Rigopoulou02,Herrero14,Herrero20}).

Recently, \citet{Efstathiou21} modelled the IR emission of local ULIRGs using the HERUS sample and found that the spheroidal host galaxy component significantly contributes to their PAH emission. The same result has also been found for LIRGs \citep{Herrero-Illana17}. Furthermore, it has also been found that
the bulk of the observed PAH emission in deeply embedded sources originates mainly in their circumnuclear regions using Spitzer data (e.g. \citealt{Marshall18} and references therein) and also subarcsecond resolution spectra (e.g. \citealt{Herrero14}). Therefore, the bulk of the PAH emission observed in the mid-IR spectrum of a galaxy containing a CON is considered relatively unobscured and is probably excited by young and hot stars (e.g. \citealt{Peeters04}), and/or old stars when present (e.g. \citealt{Kaneda08}). Fig. \ref{cartoon} depicts in a schematic the above described view of a galaxy containing a CON, which is represented by a nucleus within an optically thick, centrally heated dust shell, surrounded by a host galaxy disk (see also e.g. \citealt{Imanishi07}). Note that throughout this work we will use the term \textit{circumnuclear PAH emission} to refer to the PAH emission of the host galaxy disk which is significantly less obscured than the one originating from the compact nuclear region. 

\subsection{PAH and continuum emission}
\label{127} 
The impact of the mid-IR continuum on the circumnuclear PAH emission of U/LIRGs has subsequently been investigated with Spitzer$/$IRS spectra through the use of the EW of the 6.2~$\mu$m PAH feature (see e.g. \citealt{Armus07,Desai07,Spoon07,Imanishi07,Marshall18}). 

Taking into account the spatial scales probed by Spitzer/IRS ($\sim$4\arcsec), its resolution element includes emission from many star-forming regions and, in several cases, the whole emission of the galaxy. Consequently, PAH emission from U/LIRGs, as measured by Spitzer/IRS, is averaged over large areas of the galaxies and the PAH ratios measured from each source are expected to be roughly similar. 

Using a sample of local star-forming galaxies, \citet{Hernan-Caballero20} found a strong correlation between the 12.7/11.3~$\mu$m PAH flux ratio and the 9.7~$\mu$m silicate strength. These authors showed that the PAH EW ratio of the 12.7 and 11.2~$\mu$m PAH bands is independent of the optical depth with only a small dispersion ($\sim$5\%), suggesting that star-forming galaxies have a nearly constant intrinsic PAH flux ratio.

\begin{figure*}
\centering
\par{
\includegraphics[width=9.1cm]{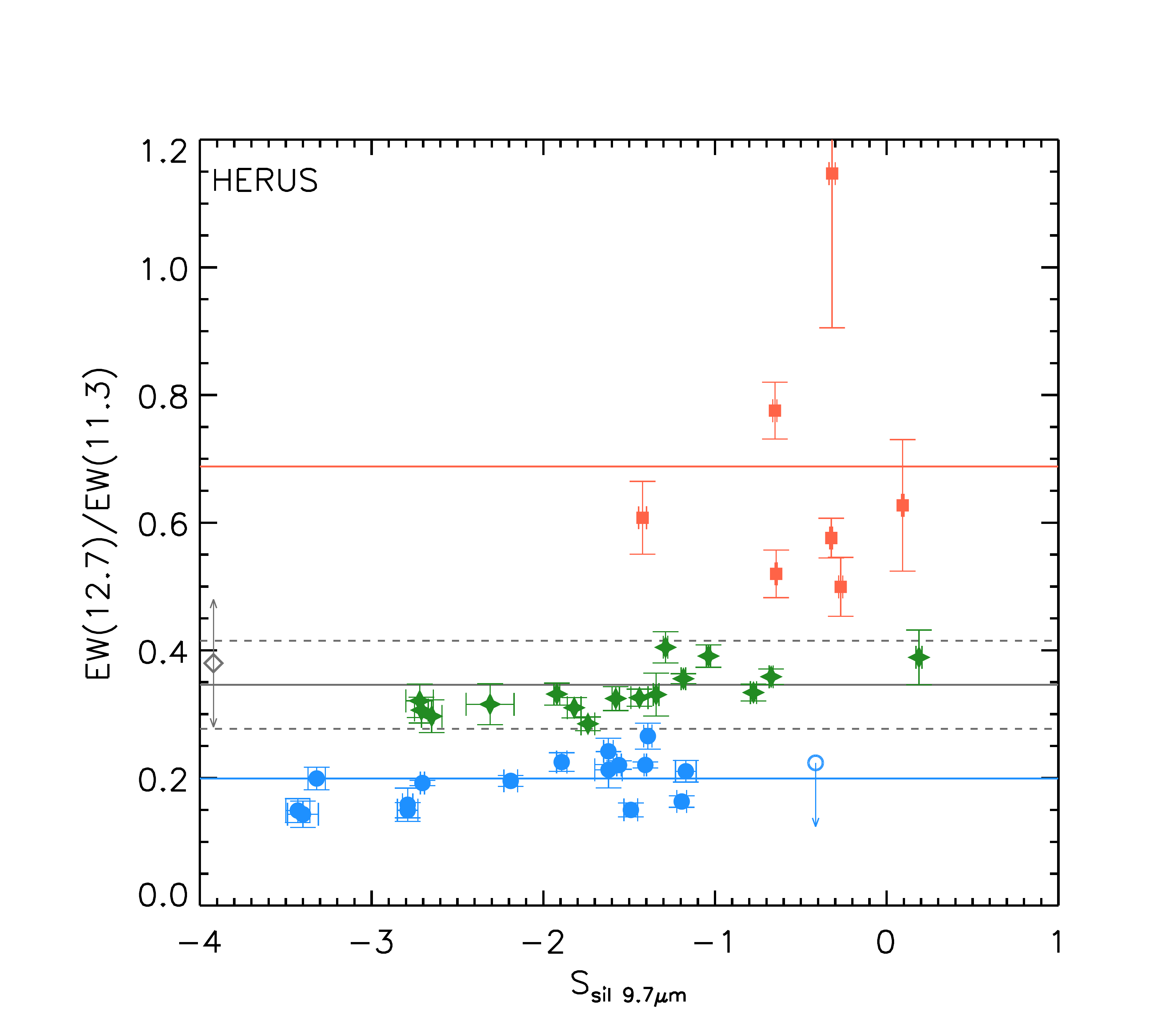}
\includegraphics[width=9.1cm]{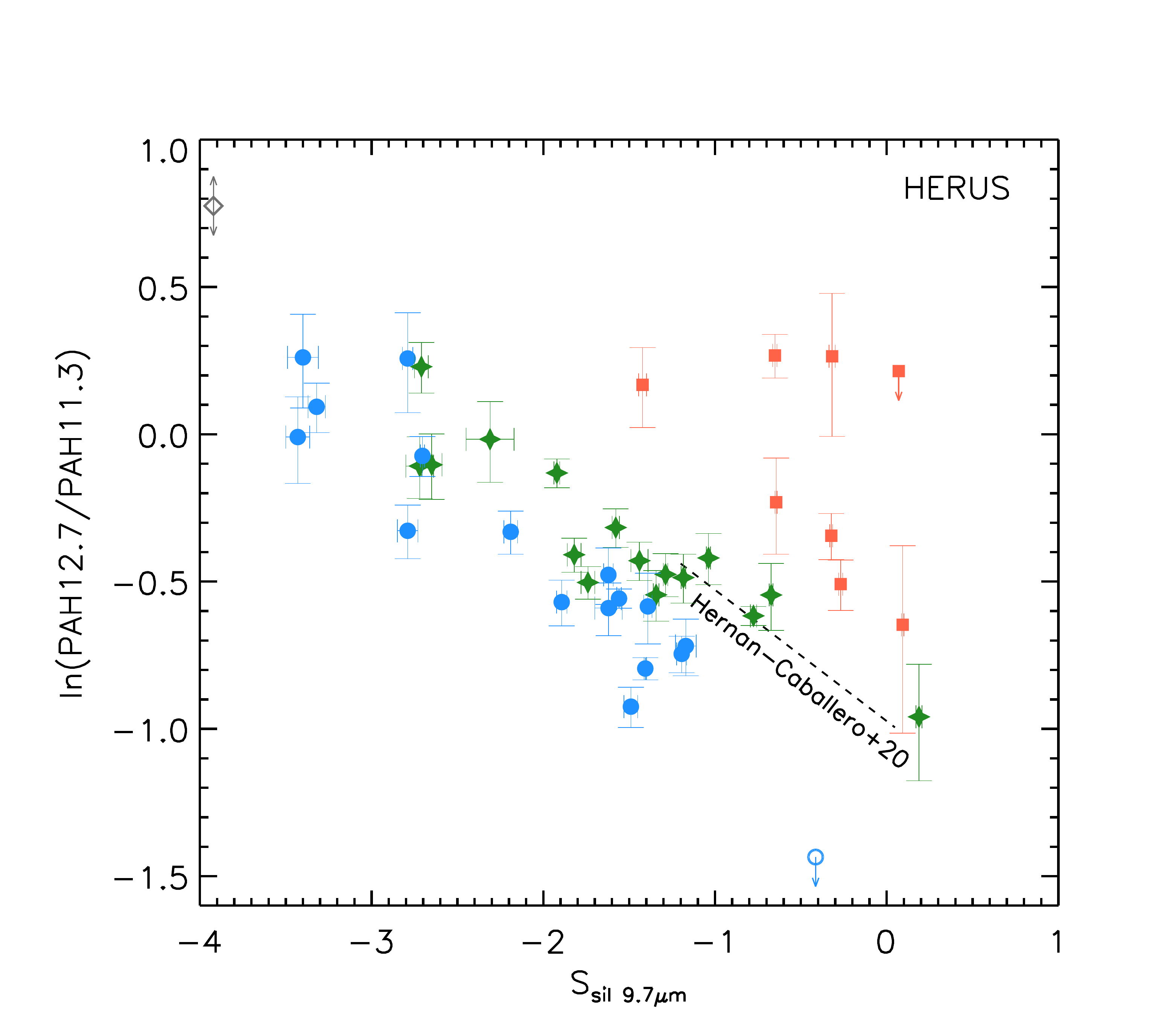}
\par}
\caption{Left panel: 12.7/11.3~$\mu$m PAH EW ratio versus the strength of the 9.7~$\mu$m silicate feature for the HERUS sample. The grey solid and dashed horizontal lines correspond to the average and 3$\sigma$ values found by \citet{Hernan-Caballero20} for star-forming galaxies. The blue and red solid horizontal lines represent the average values found for groups 2 and 3 of ULIRGs (see text). Right panel: same as the left panel but using the 12.7/11.3~$\mu$m PAH flux ratio instead of the PAH EW ratio. Green stars, blue circles and red squares correspond to groups 1, 2 and 3 of ULIRGs (see text). The black dashed line represents the correlation result from the fit of star-froming galaxies \citep{Hernan-Caballero20}. Open blue circles and open grey diamond correspond to sources with PAH EW upper limits. The silicate strength is computed as S$_{\rm Sil}=$ln(f$_{\rm peak}/$f$_{cont}$) (see Section \ref{observations}).} 

\label{ratio_diagram_eqw}
\end{figure*}

\begin{figure}
\centering
\includegraphics[width=9.5cm]{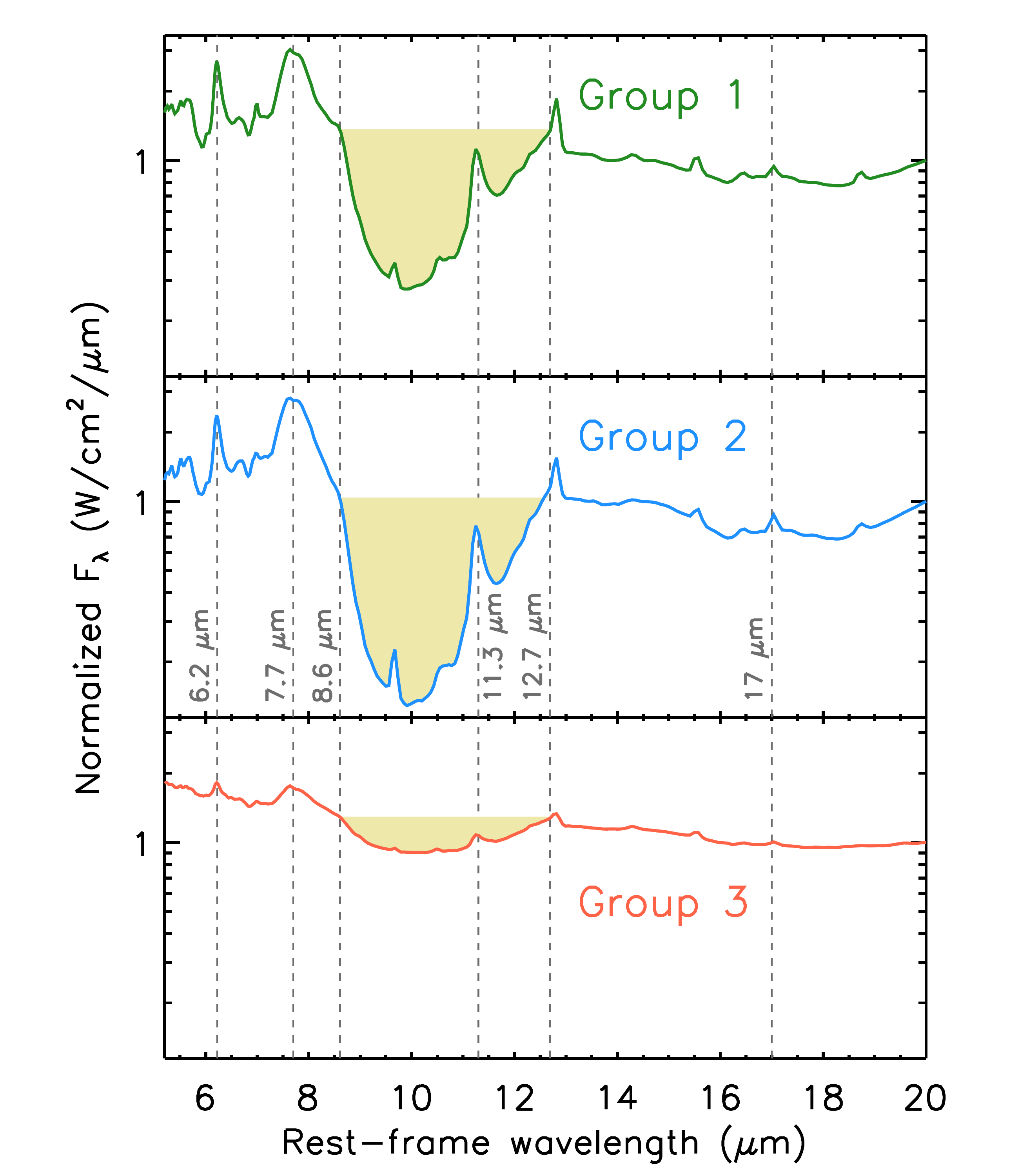}
\caption{Average spectra (normalized at 20~$\mu$m) for the three ULIRG groups defined in Section \ref{127}. The grey vertical dashed lines denote the location of the strongest PAH bands (6.2, 7.7, 8.6, 11.3, 12.7 and 17\,$\mu$m). The yellow shaded region highlight the different degrees of silicate absorptions for the various ULIRGs groups.}
\label{stacked}
\end{figure}

To investigate the relation between PAH emission and obscuration in local ULIRGs, we plot the 12.7/11.3~$\mu$m PAH EW ratio versus the silicate strength using Spitzer/IRS data for the HERUS sample (see left panel of Fig. \ref{ratio_diagram_eqw}). We find that a group of ULIRGs, hereafter group 1 (green-coloured stars), shows 12.7/11.3~$\mu$m PAH EW ratios similar to those found by \citet{Hernan-Caballero20} for star-forming galaxies (3$\sigma$; dashed black lines in left panel of Fig. \ref{ratio_diagram_eqw}). The 
remainder ULIRGs can be found in two groups, one below (group 2, blue-coloured circles) and the other above (group 3, red-coloured squares) the range occupied by star-forming galaxies (in our case group 1 ULIRGs).

To further investigate the differences between the spectral features for these groups of ULIRGs, we construct the average spectra of each group by using only those galaxies with firm PAH detections (i.e. we do not include those with upper limits). To do so, first, we correct the spectra to the rest-frame of each source and we resample them with the same wavelength grid. Then, we normalize each spectrum at 20~$\mu$m (which is a featureless part of the continuum) and, finally, we averaged the spectra of the various groups (see Fig. \ref{stacked}).

It is instructive to take a closer look at the values seen in group 2 ULIRGs. These tend to cluster around low values of the 12.7/11.3~$\mu$m PAH EW ratios (average value of $\sim$0.2; solid blue line in the left panel of Fig. \ref{ratio_diagram_eqw}) which are more than 3$\sigma$ below the mean value for star-forming galaxies. We also find that group 2 ULIRGs follow the 12.7/11.3~$\mu$m PAH flux ratio versus silicate strength relation but with an offset relative to star-forming galaxies presented in \citet{Hernan-Caballero20} (see right panel of Fig. \ref{ratio_diagram_eqw}). This indicates that in group 2 ULIRGs the silicate strength measured by Spitzer/IRS is deeper than expected from the 12.7/11.3~$\mu$m PAH flux ratio versus silicate strength relation found for normal galaxies (see Fig. \ref{stacked}). Under the assumption that the circumnuclear star-forming regions of ULIRGs have similar averaged 12.7/11.3~$\mu$m intrinsic PAH flux ratios compared to those of star-forming galaxies, the offset found in group 2 ULIRGs (described above) is consistent with an additional continuum component with very deep silicate absorption. However, group 3 of ULIRGs shows a greater dispersion in the PAH EW ratios than groups 1 and 2.

\begin{figure}
\centering
\includegraphics[width=9.5cm]{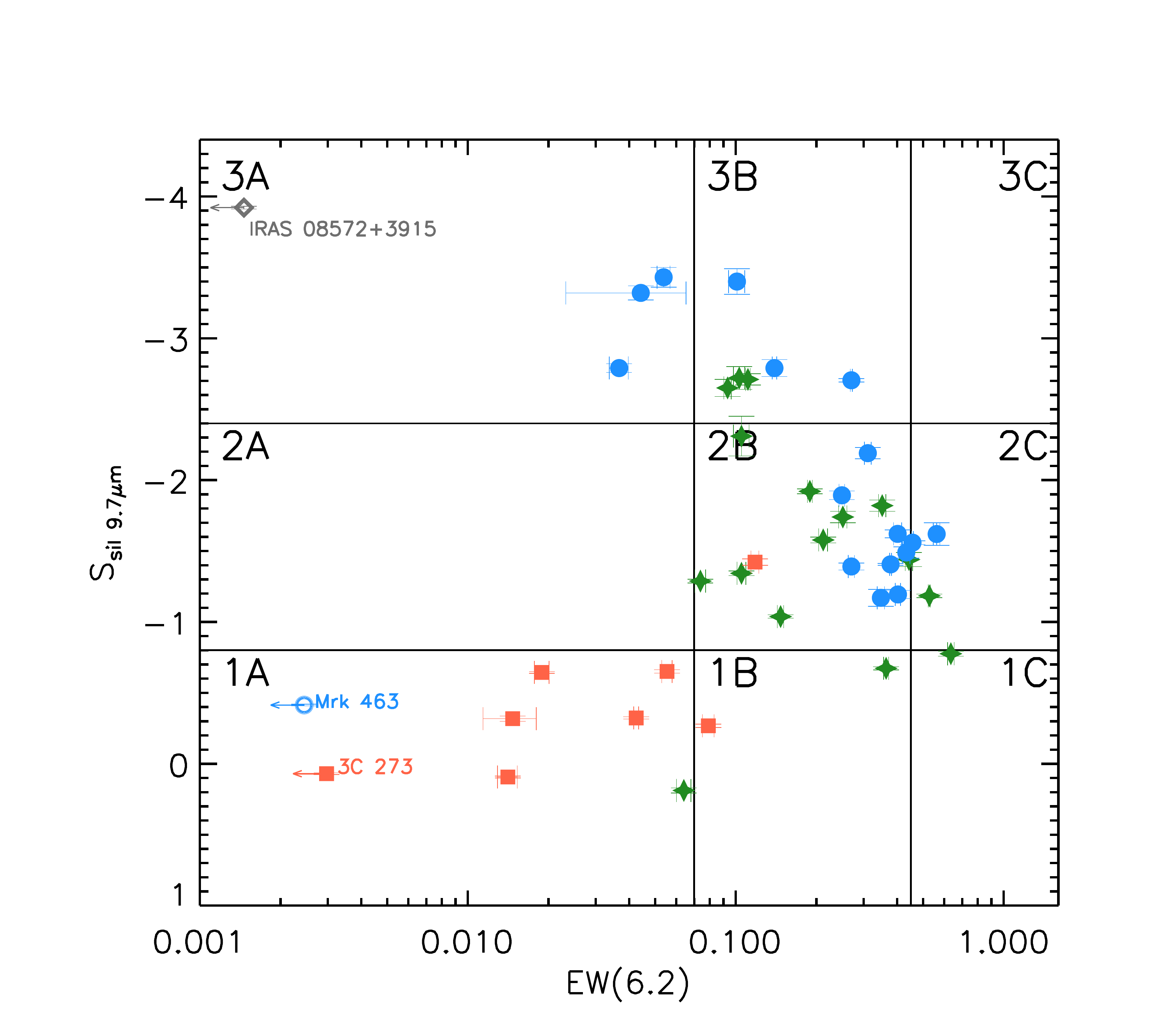}
\caption{``Fork'' diagnostic diagram (\citealt{Spoon07}) using the 6.2~$\mu$m PAH EW width versus the strength of the 9.7~$\mu$m silicate feature for the HERUS sample.
The color symbols are the same as in Fig. \ref{ratio_diagram_eqw}.} 
\label{fork}
\end{figure}

We find that the majority of ULIRGs within group 3 correspond to sources dominated by unobscured AGN (see red squares in Fig. \ref{fork}). Indeed, the sources in group 3 show weak silicate absorption or moderate silicate emission (see Figs. \ref{ratio_diagram_eqw} and \ref{stacked}), which indicates moderate to no obscuration of the nuclear mid-IR emission (i.e. a significant fraction of nuclear emission escapes through the nuclear obscurer along a viewing angle without being absorbed). As a result, the observed galaxy integrated silicate strength of group 3 ULIRGs indicates the contribution of nuclear AGN spectrum (with silicate emission) and circumnuclear PAH emission with moderate extinction. Although sources dominated by unobscured AGN show different PAH ratios, in obscured AGN the average PAH ratios measured within the Spitzer/IRS aperture are not different compared to those of star-forming galaxies (see e.g. \citealt{garcia-bernete21} and references therein). Finally, in Appendix \ref{fraction_irs} we estimate the fraction of the galaxy integrated mid-IR emission that is measured by Spitzer/IRS for the various groups of sources described above. We conclude that the fraction of the galaxy observed by Spitzer/IRS is not driving the classification of the sources used in this work.

\begin{figure*}[ht!]
\centering
\includegraphics[width=18.7cm]{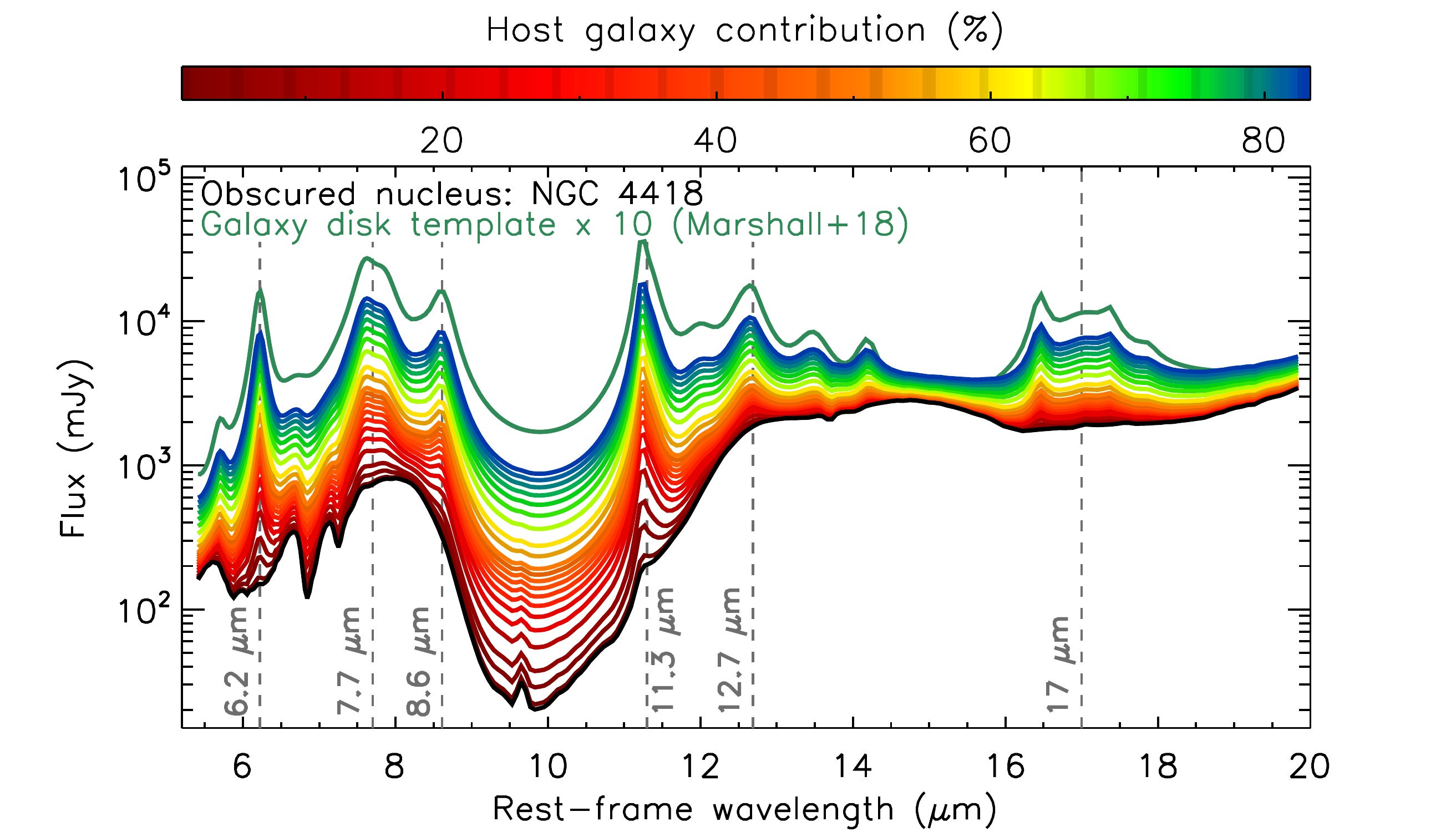}
\caption{Predicted range of mid-IR spectra of galaxies containing a CON. These spectra are generated by using different fractions of the host galaxy (represented by the host galaxy disk template from \citealt{Marshall18}; solid green line) superimposed on the Spitzer/IRS spectrum of the archetypical CON-dominated source NGC\,4418 (black solid line). The colour-code represents different fractions of the host galaxy. Dark red to blue solid lines correspond to increasing values of the host galaxy contribution with respect to the nuclear source. The grey vertical dashed lines denote the location of the strongest PAH bands (6.2, 7.7, 8.6, 11.3, 12.7 and 17\,$\mu$m).} 
\label{Synth_con}
\end{figure*}

\begin{figure*}[ht!]
\centering
\par{
\includegraphics[width=9.1cm]{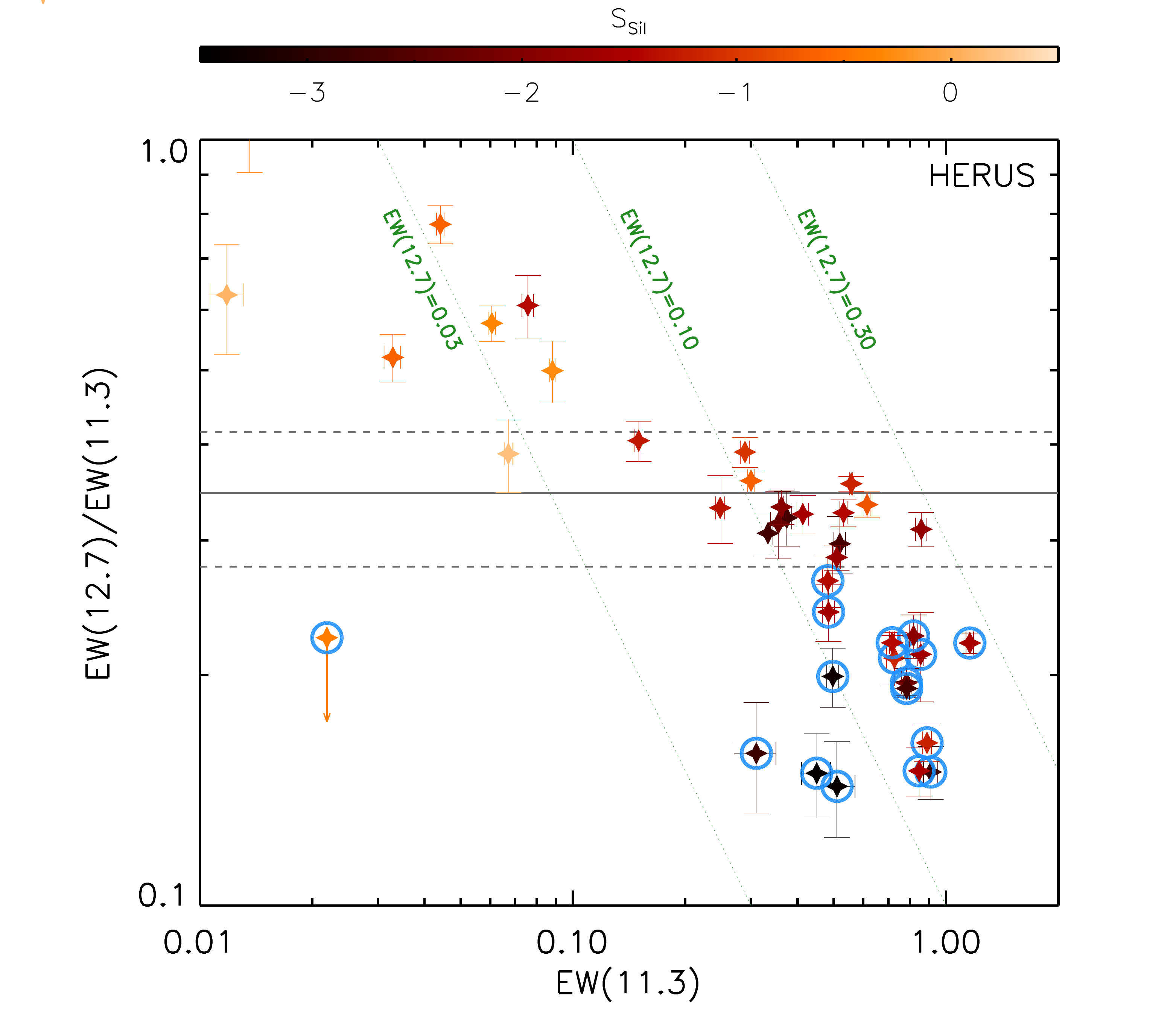}
\includegraphics[width=9.1cm]{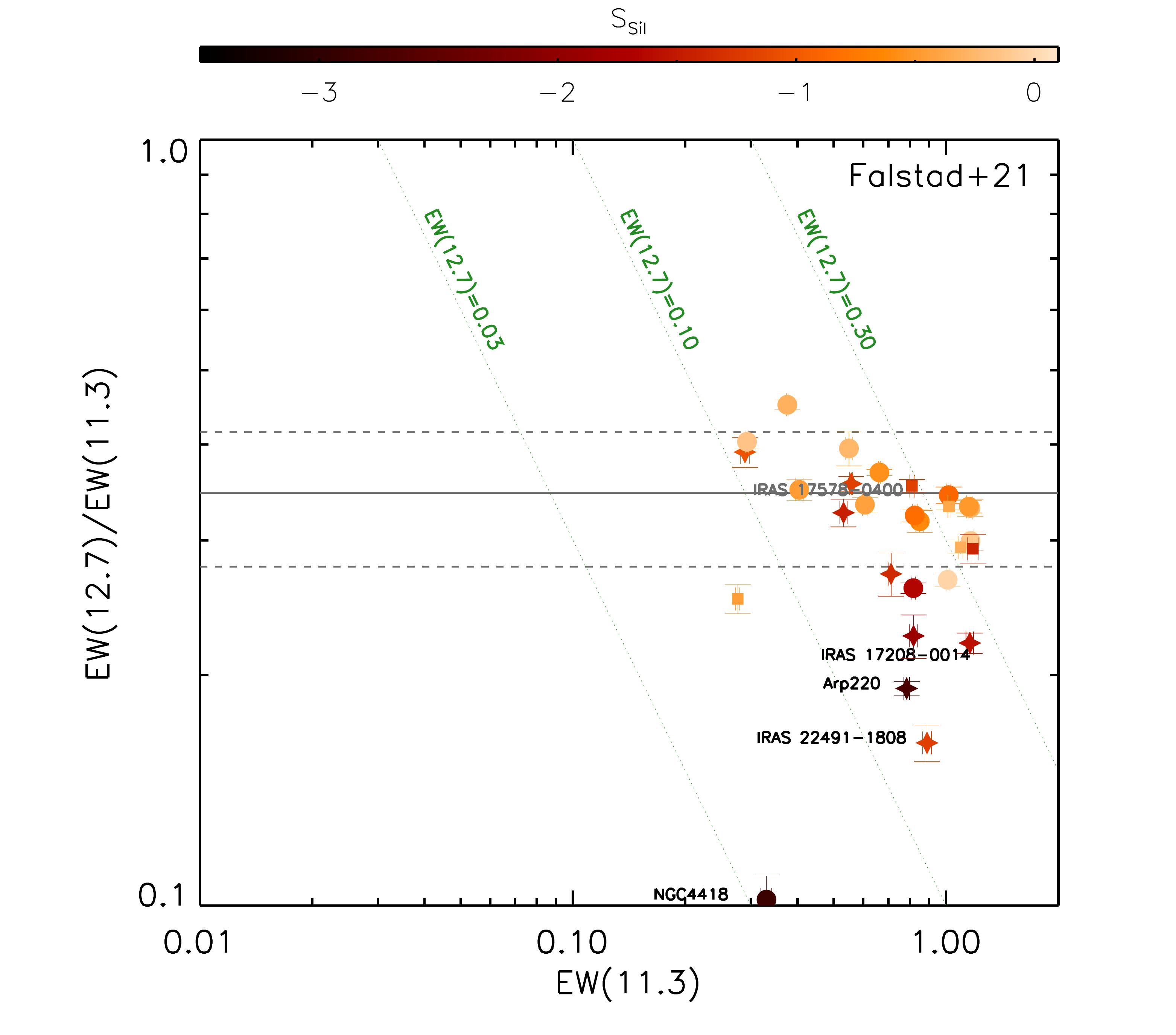}
\par}
\caption{12.7/11.3~$\mu$m PAH EW ratio versus PAH EW(11.3). Left panel: the EW plot for the HERUS sample. Open blue circles correspond to Group 2 of ULIRGs. Right panel: The EW plot for the CON-quest sample. Labelled sources correspond to those galaxies classified as CONs in \citet{Falstad21}. Color-coded stars, circles and squares represent the strength of the 9.7~$\mu$m silicate feature of ULIRGs, LIRGs and sub-LIRGs, respectively. Color-coded symbols correspond to the strength of the 9.7~$\mu$m silicate feature. The grey solid and dashed horizontal lines correspond to the average and 3$\sigma$ values found by \citet{Hernan-Caballero20} for star-forming galaxies. Finally, the green dotted lines represent constant values of the 12.7~$\mu$m PAH EW (0.03, 0.10 and 0.30).}
\label{dilution}
\end{figure*}

In deeply embedded sources, the nuclear dusty structure and the main source of the \textit{circumnuclear PAH emission} observed in the total Spitzer/IRS spectra are likely to be located in different physical regions and, thus, suffer different degrees of extinction. The EW indicates the strength of a feature compared to its underlying continuum. Therefore, assuming that the galaxy integrated PAH ratios measured from each source are roughly the same (see e.g. \citealt{Hernan-Caballero20}), any differences in the PAH EW ratios are likely not related to the emission from the circumnuclear region of U/LIRGs but rather reflect differences in the continuum that is coming from the nuclear source (see e.g. \citealt{Imanishi07}). As expected, we find an anti-correlation between PAH EW and continuum ratios, indicating that the PAH EW ratio is tracing the underlying continuum emission. We, therefore, suggest that the intrinsic shape of the IR nuclear continuum is the driving force behind the differences in the PAH EW ratios between the two groups of ULIRGs (group 1 and 2).

This idea is demonstrated in Fig. \ref{Synth_con} where we generate the IR spectra of galaxies containing CONs by using two template spectra: the first one is from NGC\,4418, a galaxy whose SED is dominated by a very deep silicate absorption feature (black solid line, representing the nuclear CON) and the other is the host galaxy template from \citet{Marshall18}, cicumnuclear PAH spectrum (green solid line representing the host galaxy providing the circumnuclear emission). By varying the contribution of the host galaxy spectrum (from dark red to blue corresponding to increasing values of the host galaxy contribution) with respect to the nuclear source we show the possible range of mid-IR spectra of the CONs. Fig. \ref{Synth_con} illustrates how the effect of the deep 9.7~$\mu$m silicate absorption feature present in CONs is particularly pronounced in their 11.3\,$\mu$m PAH feature. The low flux level of the nuclear silicate absorption band enhances the 11.3\,$\mu$m PAH feature contrast (high PAH equivalent width) compared to that of other PAH features. Consequently, the levels of dilution of the 12.7~$\mu$m (and 6.2~$\mu$m) PAH bands by the nuclear continuum is high compared to that at 11.3~$\mu$m due to the shape of the intrinsic continuum SED of the heavily obscured nucleus. 

In the left panel of Fig. \ref{dilution} we show the 12.7/11.3~$\mu$m PAH EW ratios versus the 11.3~$\mu$m PAH EW for the HERUS sample. For comparison, we also plot the CON-quest sample which includes confirmed CON sources (these are the labelled sources in the right panel of Fig. \ref{dilution}). We note that there are 7 sources in common between the HERUS and CON-quest samples (see Section \ref{sample_selection}). However, it is worth clarifying that we only use the CON-quest sample to compare the trends observed in group 2 of ULIRGs with those of confirmed CON sources. 

In general, the 12.7/11.3~$\mu$m PAH EW ratios of CON sources are below the 3$\sigma$ values found in star-forming galaxies (grey dashed line). In particular, for a given 11.3~$\mu$m PAH EW, CON sources tend to have low 12.7/11.3~$\mu$m PAH EW ratios (dotted green lines). From this plot, it is clear that the degree of dilution of the 11.3 and 12.7~$\mu$m PAH features are different in galaxies harbouring deeply obscured nuclei to those of normal star-forming galaxies.

\subsubsection{Comparisons with other dusty galaxies}
\label{sample}
\begin{figure}
\centering
\includegraphics[width=9.3cm]{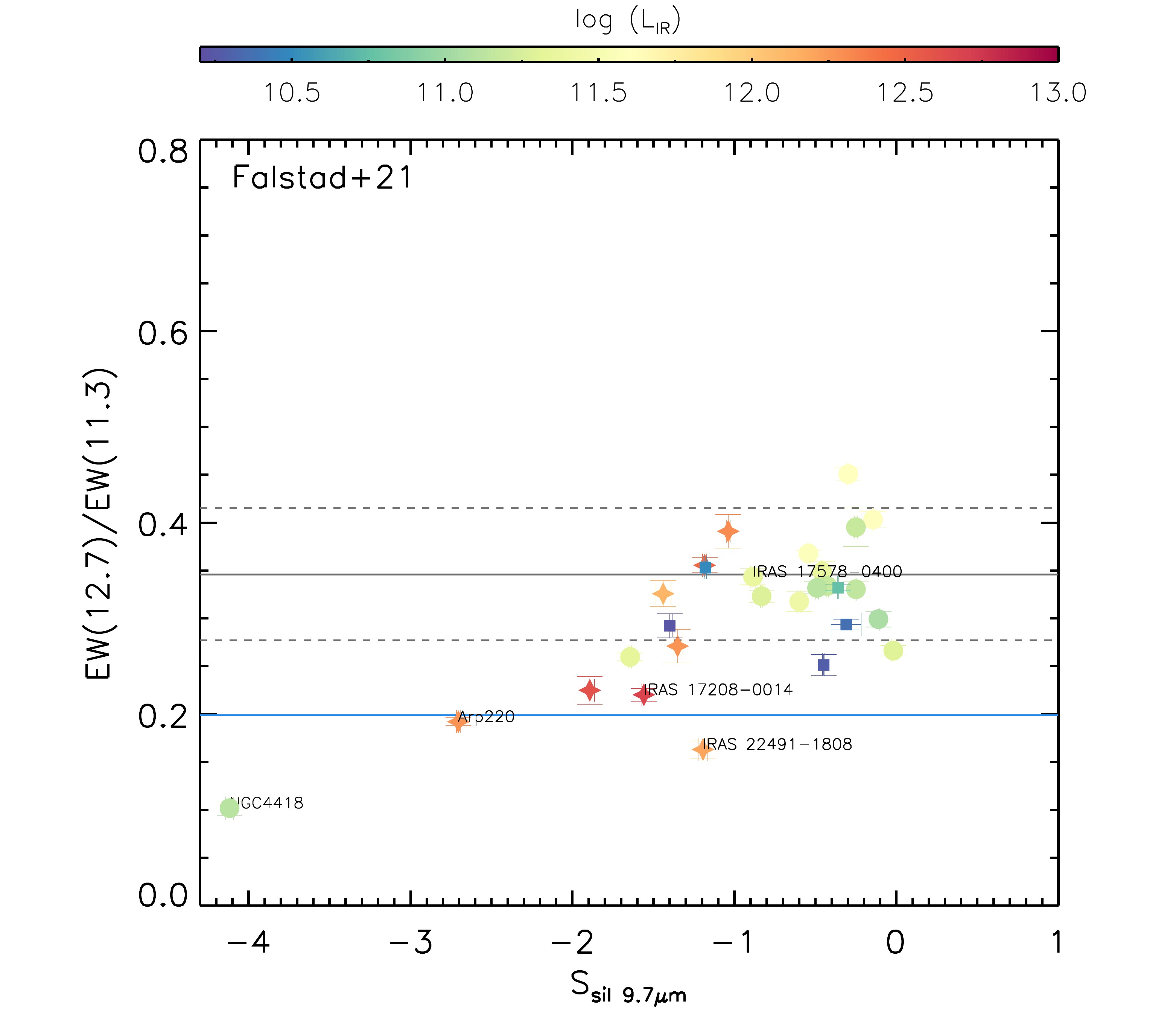}
\caption{Same as Fig. \ref{ratio_diagram_eqw} but for the CON-quest sample. Labelled sources correspond to same galaxies as in Fig. \ref{dilution} (i.e. those galaxies classified as CONs in \citealt{Falstad21}). Color-coded stars, circles and squares represent the IR luminosity of ULIRGs, LIRGs and sub-LIRGs, respectively. The grey solid and dashed horizontal lines correspond to the average and 3$\sigma$ values found by \citet{Hernan-Caballero20} for star-forming galaxies. Blue solid line corresponds to average value found for group 2 of ULIRGs (see text).} 
\label{ratio_diagram_eqwcons}
\end{figure}

To further investigate the properties of ULIRGs in group 2, which likely include galaxies with heavily obscured nuclei, we focus first on the sample of far-infrared luminous galaxies that include confirmed compact obscured nuclei (CON-quest sample; \citealt{Falstad21}). For this purpose, we plot the 12.7/11.3~$\mu$m PAH EW ratio and silicate strength of the CON-quest sample (see Fig. \ref{ratio_diagram_eqwcons}). We find that the majority of the sources classified as CONs in \citet{Falstad21} show small values of the 12.7/11.3~$\mu$m PAH EW ratio
similar to what we found earlier for ULIRGs in group 2.

However, for both CONs and group 2 ULIRGs, we find a large range in the values of the silicate strength measured by Spitzer/IRS ($\sim$4\arcsec). \citet{Levenson07} found that a foreground screen model for the obscuration of the host galaxy, cannot reproduce the 9.7~$\mu$m silicate absorption band observed in the mid-IR spectra of some ULIRGs. Given the Spitzer/IRS resolution element for our sources, the observed  9.7~$\mu$m silicate absorption band consists of a combination of various components from different physical regions. Note that any contribution from extended structures within the relatively large aperture of Spitzer/IRS can result in the filling of the silicate absorption feature (e.g. \citealt{Levenson07,gonzalez-martin13,Hatziminaoglou15}). \citet{Roche15} also found a higher flux level in the silicate minimum measured by Spitzer/IRS compared to that measured in the subarcsecond angular resolution mid-IR ground-based spectrum of the archetypal CON galaxy NGC\,4418. However, the high angular resolution spectrum of NGG\,4418 shows the same continuum flux level as the Spitzer/IRS spectrum in the range $\sim$11-13~$\mu$m (see Fig. 1 in \citealt{Roche15}). This indicates that the mid-IR slope of the continuum ($\sim$11-13~$\mu$m) is likely tracing the shape of the nuclear silicate absorption band in the integrated IR spectra of the deeply obscured nucleus NGG\,4418.

In summary, our findings suggest that the 12.7/11.3~$\mu$m PAH EW ratio (and the 11.3/12.7~$\mu$m mid-IR continuum ratio) enables us to select ULIRGs with a deep nuclear silicate absorption band even in the relatively large aperture probed by Spitzer.

\begin{figure*}
\centering
\par{
\includegraphics[width=9.1cm]{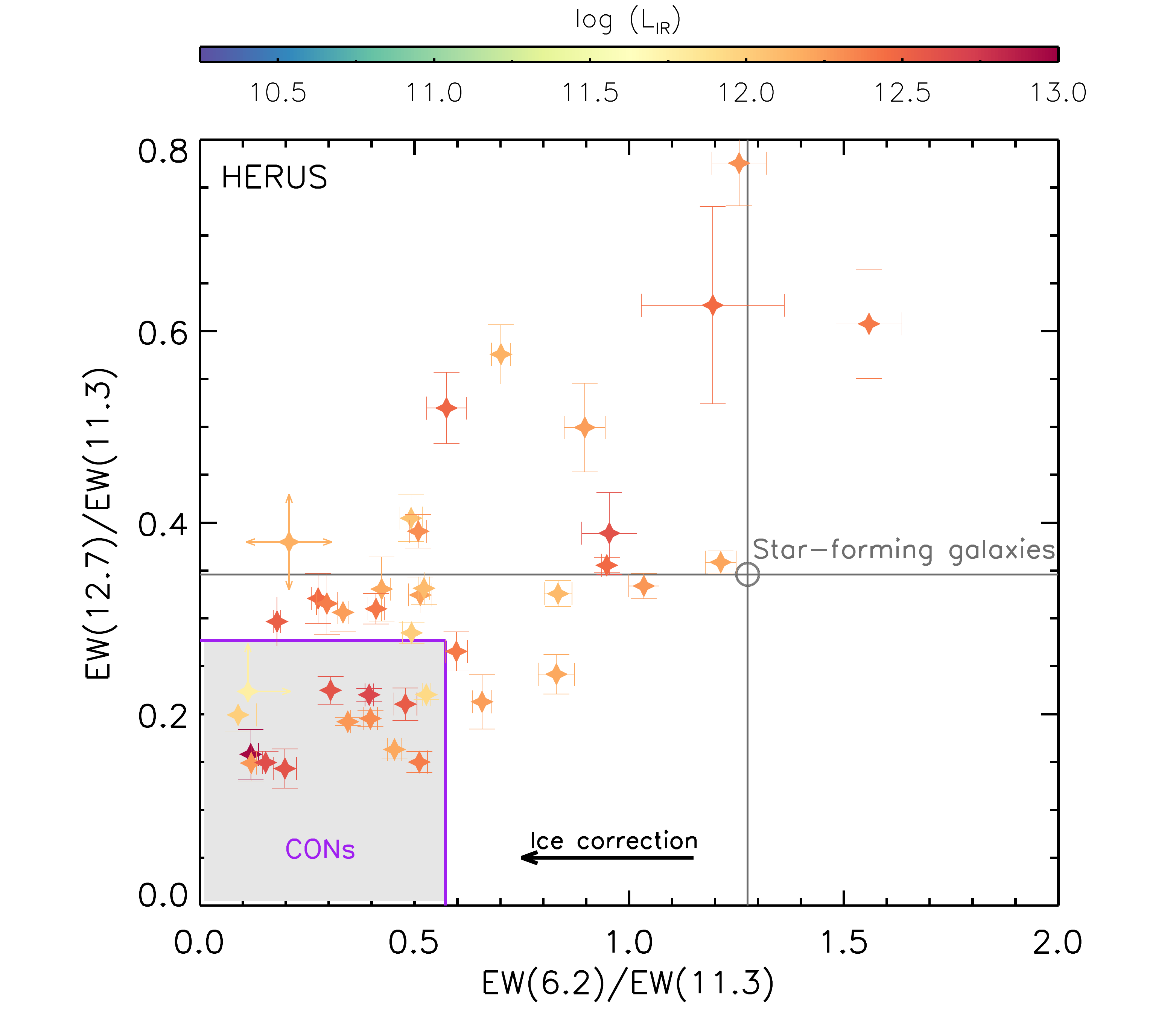}
\includegraphics[width=9.1cm]{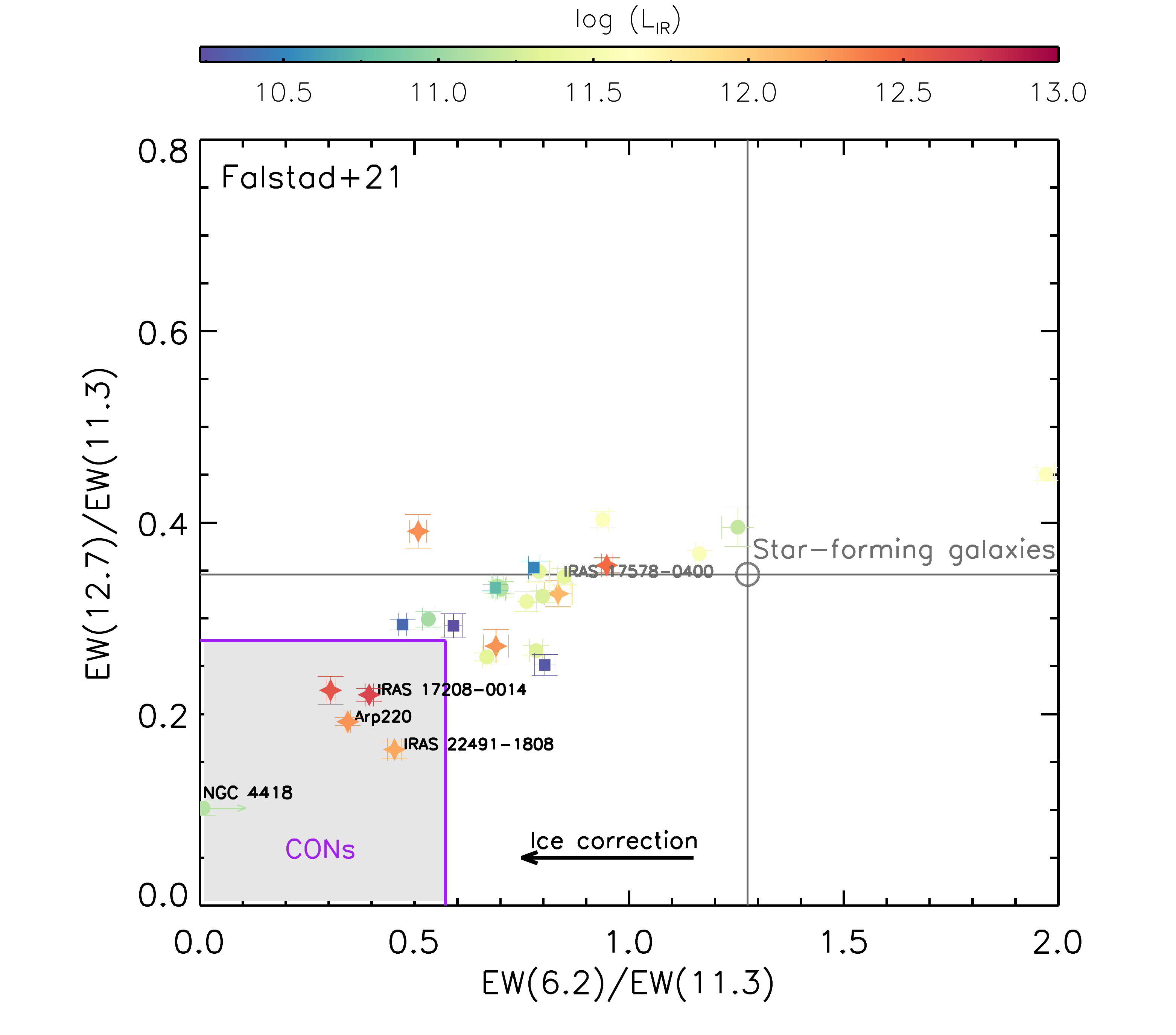}
\includegraphics[width=9.1cm]{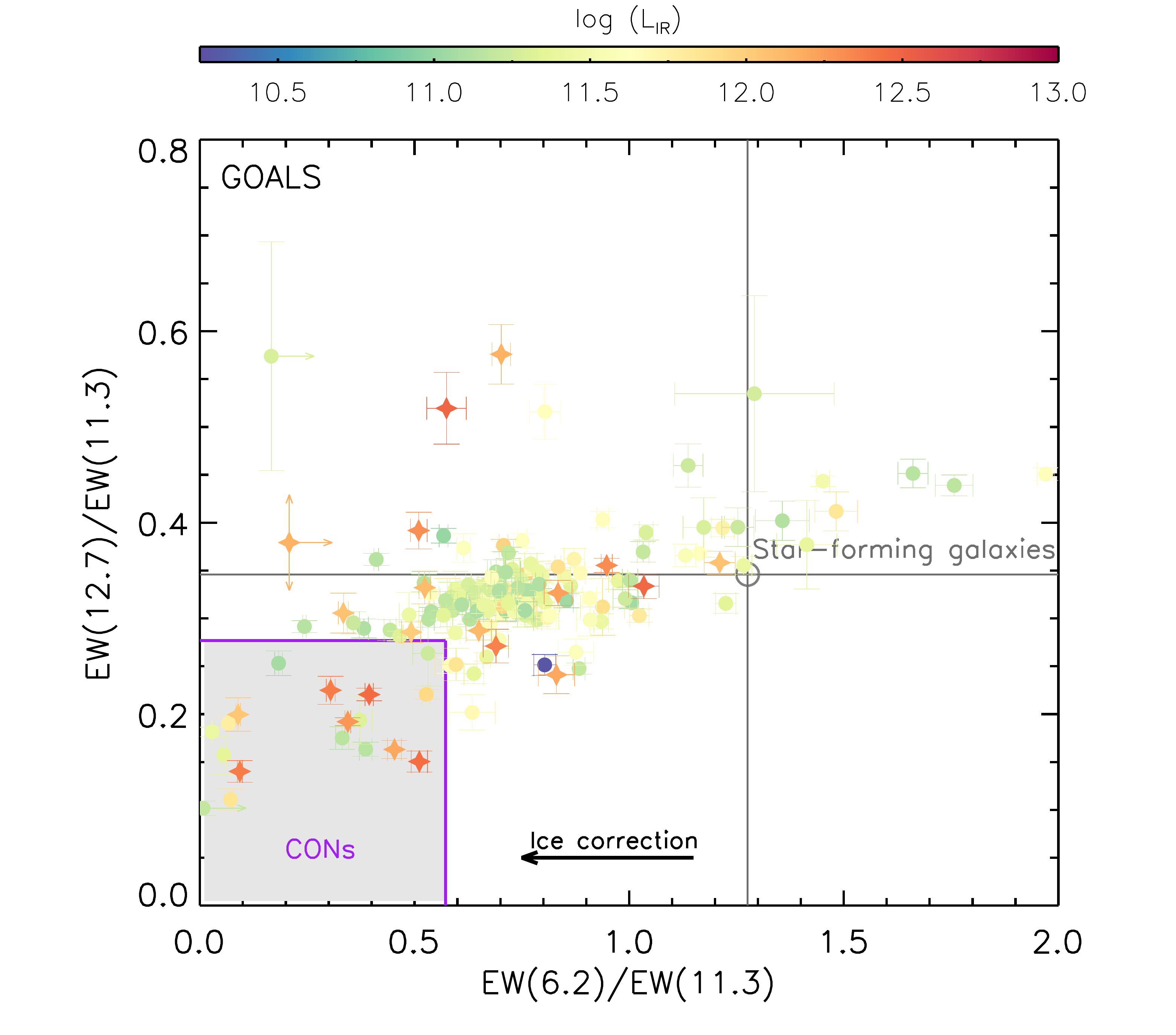}
\includegraphics[width=9.1cm]{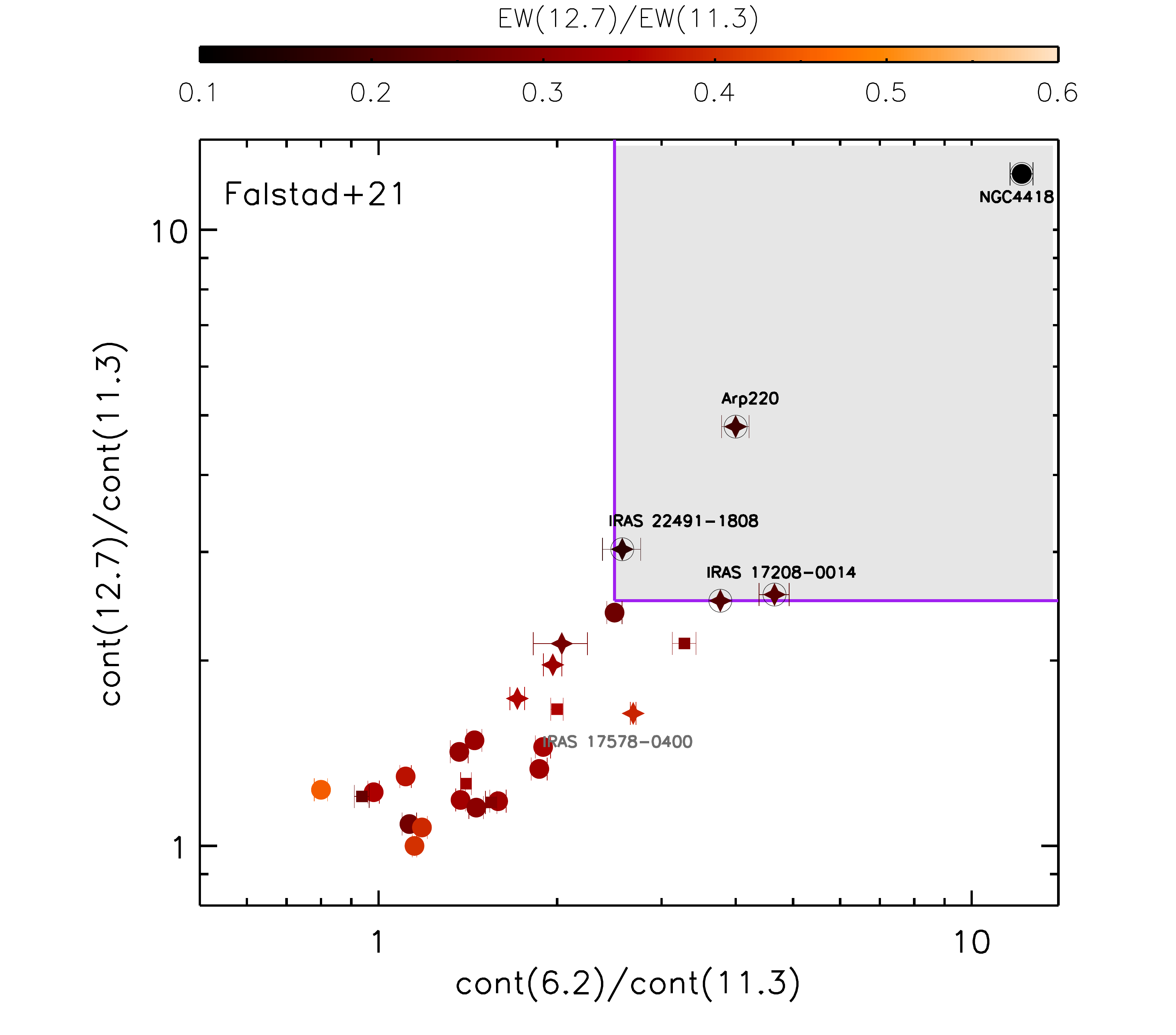}
\par}
\caption{Diagnostic diagrams for identifying heavily obscured nuclei. From top left- to bottom right-hand panels: the HERUS, CON-quest and GOALS sample. Color-coded stars, circles and squares represent the IR luminosity of ULIRGs, LIRGs and sub-LIRGs, respectively. The grey solid vertical and horizontal lines correspond to the average values for star-forming galaxies. The black arrows indicate the effect the ice absorption correction has on the 6.2/11.3~$\mu$m PAH EW ratio. The purple square denotes the region defined by the PAH EW(6.2~$\mu$m)/EW(11.3~$\mu$m)$<$ 0.572 and PAH EW(12.7~$\mu$m)/EW(11.3~$\mu$m)$<$ 0.277 (see text). Bottom right panel: the diagnostic diagram based on continuum ratios for the CON-quest sample. The purple square denotes the region defined by the continuum flux ratios: cont(11.3~$\mu$m)/cont(6.2~$\mu$m)$>$ 2.5 and cont(11.3~$\mu$m)/cont(12.7~$\mu$m)$>$ 2.5 (see text). Color-coded stars, circles and squares represent the 12.7/11.3~$\mu$m PAH EW ratio of ULIRGs, LIRGs and sub-LIRGs, respectively. Open black circles correspond to sources within the purple square in top right panel. Labelled sources correspond to those galaxies classified as CONs in \citet{Falstad21}.}
\label{EQW-EQW}
\end{figure*}

\section{A new diagnostic diagram for selecting deeply obscured galaxy nuclei}
\label{Diagnostic_diagrams}
So far the detection of the HCN--vib line has shown to be an effective method to confirm the existence of deeply obscured nuclei in local galaxies (e.g. \citealt{Sakamoto10,Imanishi13,Aalto15,Aalto19,Falstad19,Falstad21}) However, detecting CONs at high redshifts is challenging because of the faintness of the HCN--vib line. Here we investigate whether the impact of the 9.7~$\mu$m silicate absorption on the PAH EW ratios can provide a reliable alternative to unveil deeply obscured galaxy nuclei.

In Section \ref{127}, we showed that the 12.7/11.3~$\mu$m PAH EW ratios of sources with a deeply buried nucleus are different compared to those of normal star-forming galaxies (see Fig. \ref{ratio_diagram_eqwcons}). Furthermore, we found that the CON classification does not depend strongly on the silicate strength measured by Spitzer. This is because emission from the host galaxy can easily fill the silicate feature as measured by Spitzer/IRS (even if its contribution is relatively small; see Fig. \ref{Synth_con}). Due to the relatively low angular resolution afforded by Spitzer/IRS we expect some degree of contribution from the host galaxy to the total mid-IR spectra. Therefore, the integrated galaxy silicate strength alone is not a good indicator of deeply buried nuclei.

We, therefore, explore whether various PAH EW ratios and mid-IR continuum ratios could enable us to identify deeply buried nuclei. The 6.2/11.3~$\mu$m PAH EW ratio is driven by the different levels of dilution by the nuclear continuum (see Section \ref{127}). However, it is worth noting that the broad water ice absorption band peaking at $\sim$6.0~$\mu$m can affect the 6.2~$\mu$m PAH flux if it is not properly included in the mid-IR models (e.g. \citealt{Spoon07}). In that case,
the value of the EW(6.2) will increase as the continuum gets depressed by the water ice band (assuming the same intrinsic 6.2~$\mu$m PAH flux).

To mitigate this effect, the EW(6.2) values employed in our work have been corrected for the 6~$\mu$m ice absorption band (IDEOS database). 

It is also worth noting that the water ice absorption band is always weaker than the silicate absorption feature in embedded sources (see e.g. \citealt{Boogert08,Boogert11} for buried young stellar objects inside dense molecular clouds). Therefore, it is expected that deeply buried nuclei have small values of the 6.2/11.3~$\mu$m PAH EW ratio compared to normal star-forming galaxies (see Section \ref{127}). 

In addition, taking into account the geometry assumed in Fig. \ref{cartoon} for a galaxy containing a CON, the ices would cover the dusty nuclear sources and the bulk of the PAH emission would reside beyond that region. Therefore, any foreground emission will result in filling of the ice absorption feature and, consequently, any fitted ice correction can be only a lower limit of the intrinsic ice absorption band. Note that the black arrows in Fig. \ref{EQW-EQW} indicate the effect the ice absorption correction has on the 6.2/11.3~$\mu$m PAH EW ratio.

We next compare the 6.2/11.3 versus 12.7/11.3~$\mu$m PAH EW ratios of galaxies in the HERUS, CON-quest and GOALS samples, (Fig. \ref{EQW-EQW} from top left- to bottom left-hand panels). The plots reveal that there are marked differences in the PAH EW values of sources containing deeply buried nuclei and those of star-forming galaxies. Using the CON-quest sample as our ``testbed'' sample for CON sources, our method selects 4 out of the 5 sources already classified as CONs by \citet{Falstad21} (5 of 6 by including the prototypical CON IC\,860) by using the purple box showed in Fig. \ref{EQW-EQW}. To define this box, we use the 3$\sigma$ value from the average 12.7/11.3 PAH EW and 2$\sigma$ value from the average 6.2/11.3 PAH EW of the star-forming galaxy sample used in \citet{Hernan-Caballero20}. Note that two LIRGs classified as CONs by \citet{Falstad21} (namely Zw\,049.057 and ESO\,320-G030) are not included in this work since they have not been observed in the Spitzer/IRS staring mode. Fig. \ref{EQW-EQW} demonstrates how the presence of a heavily obscured nucleus can be unveiled using the PAH EW(6.2~$\mu$m)/EW(11.3~$\mu$m)$<$ 0.572 and PAH EW(12.7~$\mu$m)/EW(11.3~$\mu$m)$<$ 0.277 criteria\footnote{Note that these criteria can vary depending on the employed method to fit the mid-IR continuum and the dust emission features.}. In essence, what these ratios measure is the different degree of dilution of the 11.3~$\mu$m PAH compared with that of the 12.7~$\mu$m (and 6.2~$\mu$m) PAH due to the silicate absorption feature.

Finally, in order to investigate the existence of deeply obscured nuclei in sources with non-detection in the PAH bands we introduce an additional diagnostic tool based on the continuum fluxes at 6.2, 11.3 and 12.7~$\mu$m (bottom right panel of Fig. \ref{EQW-EQW}). Using the CON-quest sample we confirm that the diagnostic plot based on 11.3$/$12.7 and 11.3$/$6.2~$\mu$m continuum flux ratios is also capable of identifying CONs. For the continuum ratios, we define the region for selecting CON-candidates by ensuring the selection of the sources classified with the PAH EW criteria. In particular, we use the following continuum flux ratios: cont(11.3~$\mu$m)/cont(6.2~$\mu$m)$>$ 2.5 and cont(11.3~$\mu$m)/cont(12.7~$\mu$m)$>$ 2.5 (see purple box in bottom right panel of Fig. \ref{EQW-EQW}). Although the continuum ratios method can be useful when the 6.2, 11.3 and/or 12.7~$\mu$m PAH bands are not detected, the method does present some limitations. The PAH EW ratio allows discarding sources with deep silicate absorption due to foreground absorbers since they are affecting both the continuum and circumnuclear PAH emission equally. Thus, the foreground extinction is cancelled out in the PAH EW ratio. However, the continuum ratios are affected both by the extinction coming from the host galaxy and from the nuclear region. Note that the former can be the dominant one in edge-on galaxies (e.g. \citealt{Goulding12}).

\subsection{Identifying CONs using mid-IR spectroscopy}
By applying the PAH EW criteria described above we were able to identify as CON candidates 13 ULIRGs from the HERUS sample, and 10 LIRGs and 8 ULIRGs from the GOALS sample. Note that 7 of the 8 ULIRGs identified as CON candidates are common between the GOALS and the HERUS samples (see Section \ref{sample_selection} for more details on the overlap between the two samples). Therefore, our diagnostic method identifies 30\% ULIRGs and 7\% LIRGs from the HERUS and GOALS samples. It is worth noting that our method identifies IRAS\,14348-1447, as a CON candidate although the galaxy has a $\Sigma_{\rm HCN-vib} $<$ 1~{\rm L}_{\sun}$ pc$^{-2}$ and therefore is not formally identified as CON in \citealt{Falstad21}.

The percentage of ULIRGs that are identified as CON candidates by our method agrees well with those reported by \citet{Falstad21}. However, our method recovers a lower number of LIRG CON candidates compared to \citet{Falstad21} (7\% vs 21\%). A possible explanation for the lower detection rate of CONs in LIRGs could be the higher levels of contribution from the host galaxy emission to the total mid-IR spectrum as probed by Spitzer (see Fig. \ref{Synth_con}). This is related to the fact that the mid-IR continuum emitting regions of LIRGs are not as compact as those in ULIRGs (e.g. \citealt{Diaz-Santos08}, \citealt{Diaz-Santos10}; see also Appendix \ref{fraction_irs}). Thus, considering the angular resolution of Spitzer/IRS ($\sim$4\arcsec), we expect a higher contribution from the host galaxy to the mid-IR spectra of LIRGs. Finally, we find no CON candidate in the sub-LIRGs sample (L$_{\rm IR}<$10$^{11}$ L$_{\sun}$), in agreement with the results reported by \citet{Falstad21}. 

Considering our lower detection rate (i.e. 7\%) and taking into account the U/LIRG number density in the local Universe ($\sim$6$\times$10$^{-5}$ Mpc$^{-3}$; \citealt{Sanders03}), we infer a CON density $\gtrsim$4.1$\times$10$^{-6}$ Mpc$^{-3}$ (see also \citealt{Falstad21}). The higher detection rate in ULIRGs can be related to the lower dilution of the CON emission by the host galaxy in high luminous sources. 

Finally, we note that IRAS\,17578-0400, which is classified as a CON in \citet{Falstad21}, is not selected by the PAH EW (or continuum) ratios. This LIRG has a highly inclined galaxy disk (i$\sim$80$^\circ$), which is part of the interacting system in an early stage of merging \citep{Stierwalt13}. Because of the low spatial resolution afforded by Spitzer/IRS, together with the fact that LIRGs have relatively extended mid-IR emission, it is likely that the galaxy integrated mid-IR spectrum of IRAS\,17578-0400 is dominated by circumnuclear emission. Thus, higher spatial resolution data is needed to properly isolate the nuclear source. Fig. \ref{Synth_con} illustrates the impact of the host galaxy contribution to the total IR emission of galaxies containing CONs (see also Section \ref{sample}).

\begin{figure}[ht!]
\centering
\par{
\includegraphics[width=9.02cm]{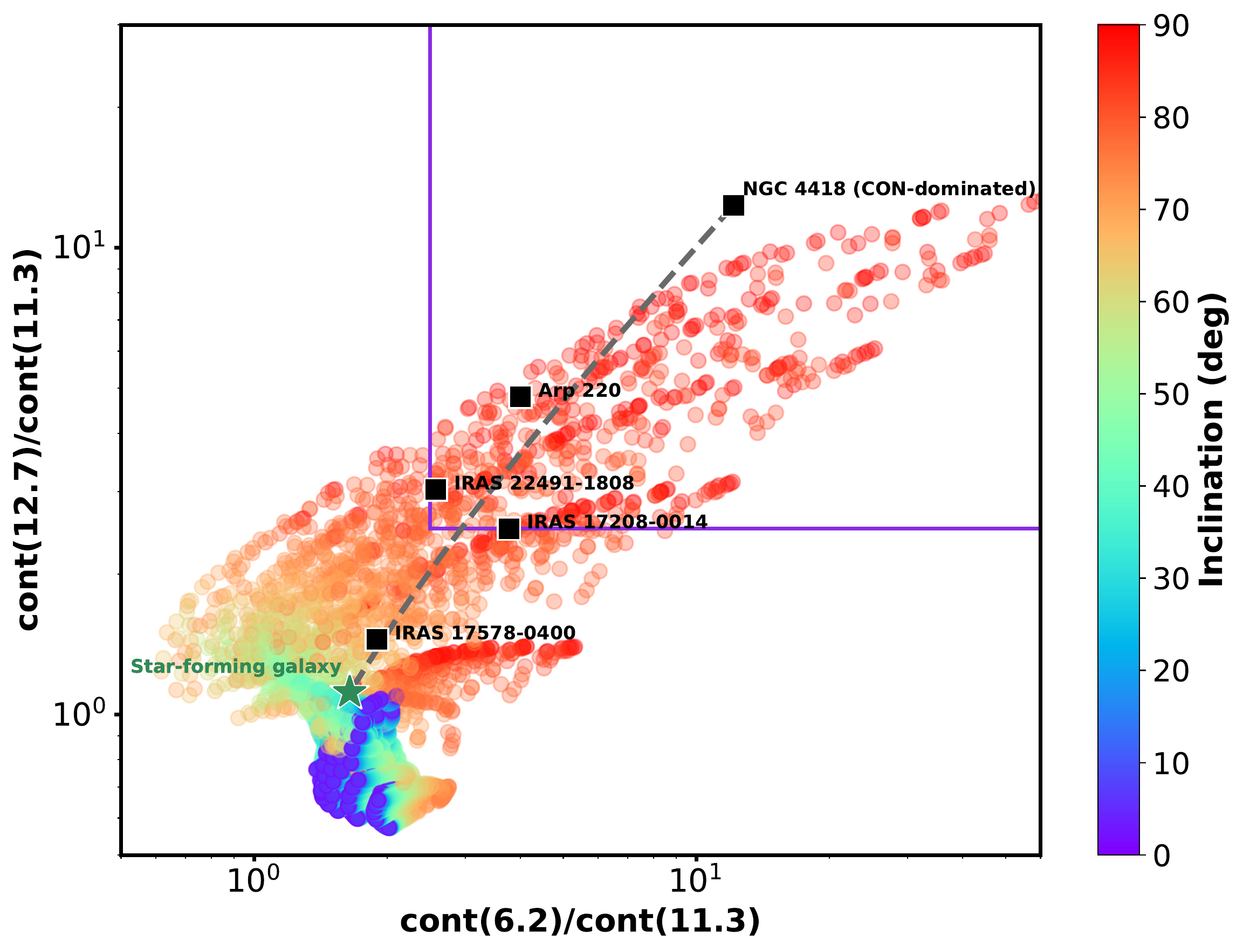}
\includegraphics[width=9.29cm]{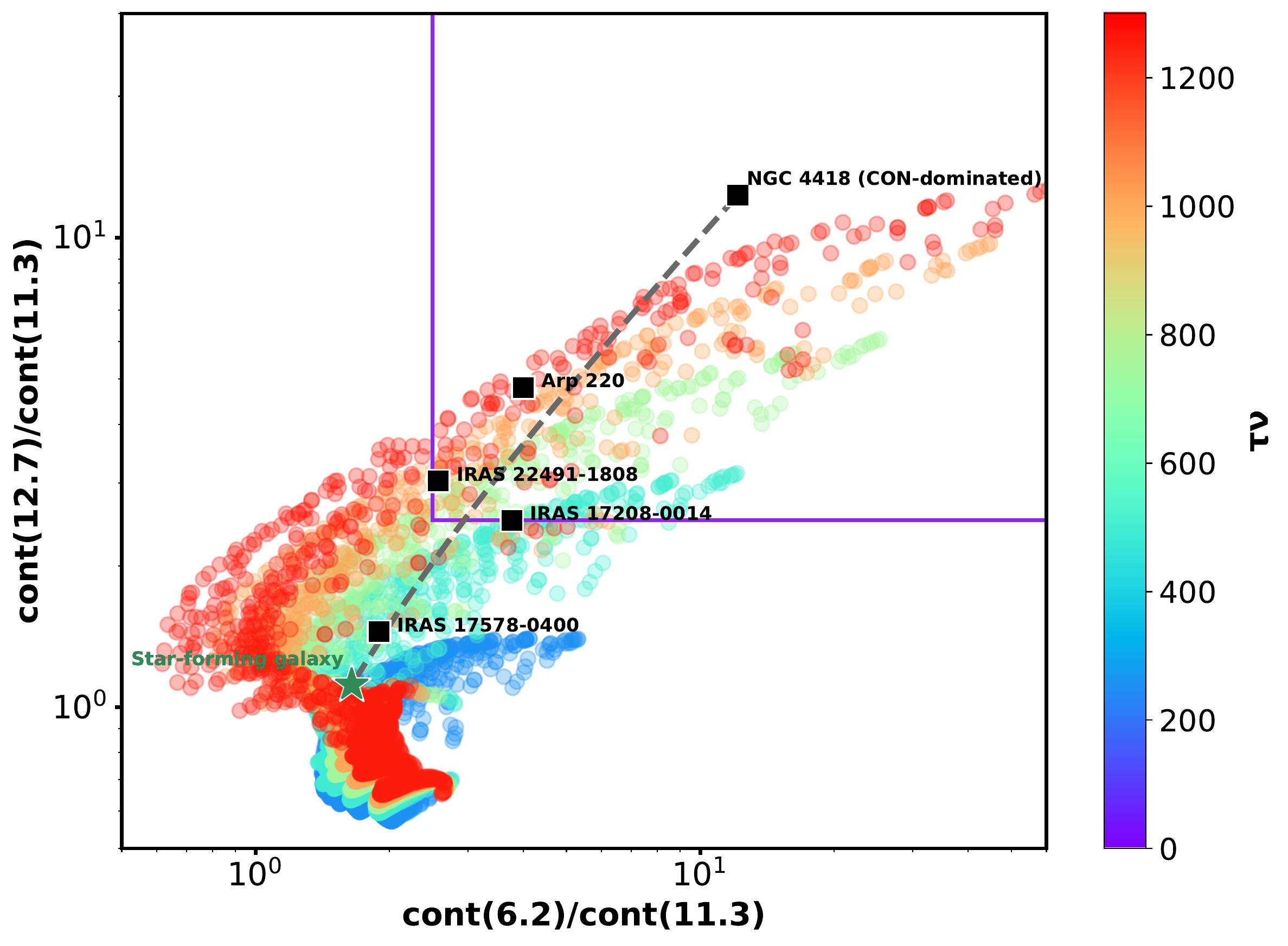}
\par}
\caption{Comparison of the observed continuum ratios seen in CONs (12.7/11.3~$\mu$m vs 6.2/11.3~$\mu$m) to those computed from the smooth CYGNUS models by \citet{Efstathiou95,Efstathiou21}. The black squares correspond to the observed values of the CONs. The green star represents a normal star-forming galaxy. Color-coded circle symbols represent the torus inclination angle and equatorial optical depth ($\tau_{1000\,\AA}$) in top and bottom panels, respectively. The purple square denotes the region where CON-dominated sources are located and defined by the continuum flux ratios: cont(11.3~$\mu$m)/cont(6.2~$\mu$m)$>$ 2.5 and cont(11.3~$\mu$m)/cont(12.7~$\mu$m)$>$ 2.5. The dashed grey line correspond to a track of simulated spectra for galaxies containing a CON varying contributions of the host galaxy as presented in Fig. \ref{Synth_con}.}
\label{torus_fig}
\end{figure}

\begin{figure*}[ht!]
\centering
\par{
\includegraphics[width=9.0cm]{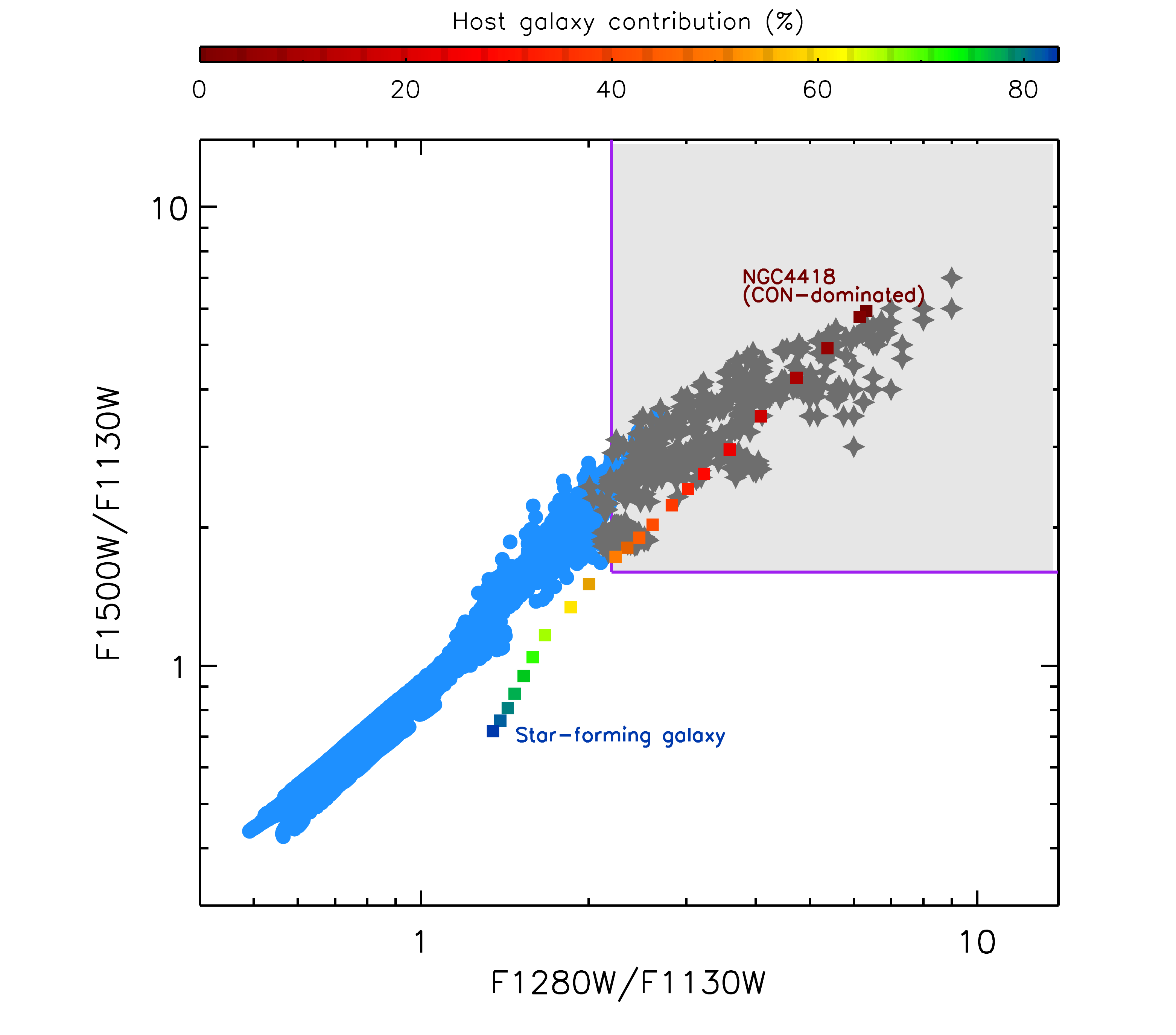}
\includegraphics[width=9.0cm]{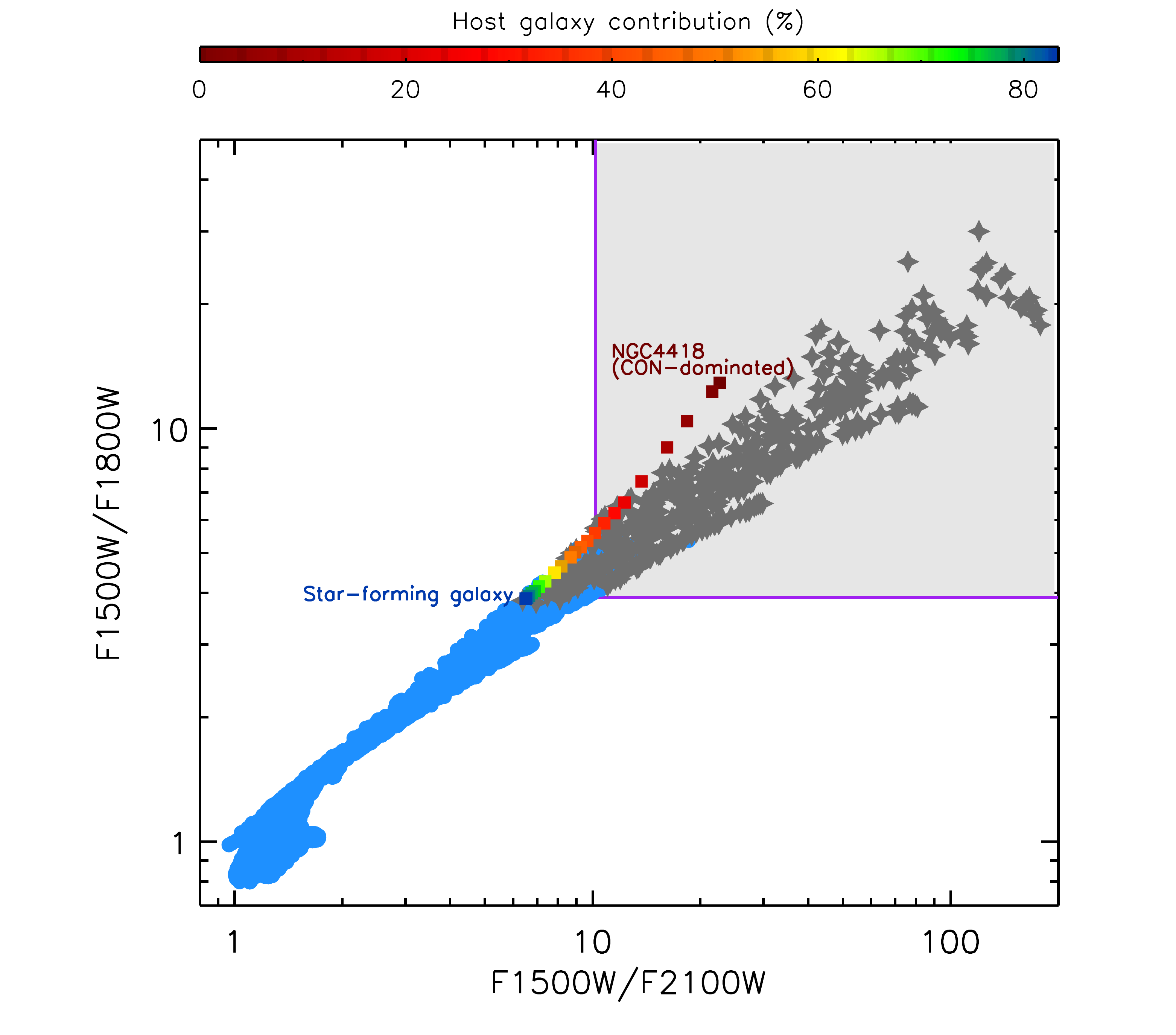}
\par}
\caption{Colour--colour diagrams using JWST filters. Left panel: smooth CYGNUS torus model SEDs at z$=$0. Right panel: smooth CYGNUS torus model SEDs at z$=$1. Blue circles correspond to the broad-band filter ratio for all the SEDs of smooth CYGNUS torus models. Grey stars represent those ratios of the model SEDs of deeply obscured nuclei (classified based on the continuum ratio criteria presented in Section \ref{Diagnostic_diagrams}). The purple square denotes the region where CON-dominated sources are located (see Section \ref{filter}). Color coded squares correspond the host galaxy contribution of the track of simulated spectra for a galaxy containing a CON showed in Fig. \ref{Synth_con}.}
\label{colorcolor_fig}
\end{figure*}

\section{Prospects for identifying compact obscured nuclei with the James Webb Space Telescope}
\label{filter}

The recent launch of the James Webb Space Telescope (JWST) opens an alternative route to identifying CONs by exploiting the presence of the 9.7~$\mu$m silicate absorption and its impact on the IR emission. The unprecedented combination of sensitivity and high spatial resolution at the mid-IR range will allow us to investigate the most distant and faintest galaxies. Therefore, it is of interest to examine the possible use of broad-band JWST filters to unveil compact obscured nuclei in local and high redshift sources.

To produce a very deep nuclear silicate absorption feature, the dusty structure must be in a centrally heated dust geometry with a temperature gradient and have an extremely high dust coverage (i.e. optical depth) due to cold absorbing material, at least towards our line-of-sight, since any warm emitting material that has a significantly lower obscuration would result in the filling of the silicate absorption feature (e.g. \citealt{Smith89,Levenson07,gonzalez-martin13,Roche15}). Therefore, it is expected that torus models with a high covering factor and smooth dust distribution could be well suited to reproduce the observed continuum ratios of deeply obscured galaxy nuclei (see e.g. \citealt{Levenson07}). 

In Fig. \ref{torus_fig} we compare the observed continuum ratios (12.7/11.3~$\mu$m vs 6.2/11.3~$\mu$m) with those obtained from the CYGNUS smooth torus models by \citet{Efstathiou95} \& \citet{Efstathiou21}. We use the same continuum ratios (i.e. 12.7/11.3~$\mu$m vs 6.2/11.3~$\mu$m) as those presented in Section \ref{Diagnostic_diagrams} (Fig. \ref{EQW-EQW}). We find that the observed continuum ratios of CON-dominated sources can be explained by assuming a tapered disc geometry and smooth torus models. This suggests that the nuclear dusty structure of deeply obscured galaxy nuclei has, indeed, an extremely high dust coverage. In addition, the best suited torus SEDs for reproducing the observed continuum ratios of CON-dominated sources tend to have relatively high equatorial optical depths ($\tau_{1000\,\AA}$) and almost edge-on values of the torus inclination angle (see Fig. \ref{torus_fig}).

In order to test the ability of JWST mid-infrared instrument (MIRI) imager to isolate deeply obscured nuclei, we use the predictions of the smooth CYGNUS torus models (see Fig. \ref{torus_fig}) and combinations of the various JWST filter transmission curves. F1300W filter mainly tracks the 11.3\,$\mu$m PAH feature and part of the 9.7\,$\mu$m silicate band (also F1000W), and F1280W covers the 12.7\,$\mu$m PAH feature and [Ne\,II] emission line. F1800W samples the 18\,$\mu$m silicate band, whereas F1500W, F2100W and F2500W sample the continuum and weaker features. To select those SEDs that are representative of CON-dominated sources (i.e. relatively low host galaxy contribution) we use the continuum ratio criteria presented in Section \ref{Diagnostic_diagrams}.

In the left panel of Fig. \ref{colorcolor_fig} we present a JWST colour--colour diagram based on ratios of the F1130W, F1280W and F1500W filters. We define a region (purple box) in this diagram for maximizing the selection of CON-dominated sources (i.e. F1280W/F1130W$>$ 2.2 and F1500W/F1130W$>$ 1.6). Using this diagram at z$=$0, we recover $\sim$95\% of the CONs (i.e. model SEDs representing CON sources) with $\sim$8\% of contamination from non-CON SEDs. However, even in CON-dominated sources, a small contribution of the host galaxy is expected. Thus, for comparison, we also plot the expected values of the predicted spectra of galaxies containing a CON showed in Fig. \ref{Synth_con} (color squares in Fig. \ref{colorcolor_fig}). We find that this diagram is effective for host galaxy contribution to the mid-IR spectrum $\lesssim$50\%. Similar results are obtained using also combinations of F1000W, F1800W and F2100W filters. Finally, we investigate the combination of additional JWST filters which will enable us to select CONs at higher redshifts. The right panel of Fig. \ref{colorcolor_fig} shows that the F1500W/F2100W-- F1500W/1800W diagram is effective at selecting CONs at z$=$1. Using the latter diagram, we recover $\sim$91\% of the CONs (see purple region of the right panel of Fig. \ref{colorcolor_fig}; F1500W/F2100W$>$10.2 and F1500W/1800W$>$3.9) with $\sim$6\% of contamination from non-CON SEDs. We find similar results for z$=$0.5 and 1.5 using the F2100W/F1500W--F1800W/F1500W and F1800W/F2500W--F2100W/F2500W diagrams, respectively.  Given the effectiveness of the mid-IR color–color diagrams for unveling CON-dominated sources, JWST will offer a unique opportunity to push the identification of CONs beyond the local Universe. Furthermore, the combination of JWST's high spatial resolution and PSF stability will result in minimising the impact of the host galaxy in the measured mid-IR emission of CONs. In a forthcoming paper, we will investigate JWST/MIRI simulations of galaxies containing a CON to further evaluate the ability of JWST to correctly isolate the nuclear emission from the circumnuclear galaxy emission in these sources.

\section{Conclusions}
\label{conclusions}
We have presented a method for identifying compact obscured nuclei that relies on the effect of the silicate absorption band on the EW of the PAH emission. The method was applied to Spitzer/IRS spectra of a representative sample of local U/LIRGs. The main results of our study are as follows.\\

   \begin{enumerate}
   
\item The intrinsic shape of the IR nuclear continuum is the driving force behind the differences in the 6.2/11.3 and 12.7/11.3\,$\mu$m PAH EW ratios in U/LIRGs. We find a strong anticorrelation between the galaxy integrated PAH EW ratio and the ratio of the underlying continuum of the PAH features.\\

\item In compact obscured nuclei we find that the effect of the 9.7\,$\mu$m silicate absorption band is particularly pronounced in the EW of the 11.3\,$\mu$m PAH feature. The low flux level of the nuclear silicate absorption band enhances the 11.3\,$\mu$m PAH feature contrast (high PAH equivalent width) compared to that of the other PAH features (i.e. 6.2 and 12.7~$\mu$m PAH bands).\\

\item Using
the 12.7/11.3~$\mu$m PAH equivalent width ratio (and the 11.3/12.7~$\mu$m mid-IR continuum ratio) we are able to select sources with a deep nuclear silicate absorption band even in the relatively large aperture probed by Spitzer ($\sim$4\arcsec). \\

\item We introduce a method to 
identify heavily obscured nuclei 
based on PAH equivalent width ratios. We demonstrate that the use of PAH EW(6.2~$\mu$m)/EW(11.3~$\mu$m)$<$ 0.572 and PAH EW(12.7~$\mu$m)/EW(11.3~$\mu$m)$<$ 0.277 is able to select CON sources in samples of local U/LIRGs. Note that these criteria can vary depending on the employed method to fit the mid-IR continuum and the dust emission features. We extend the technique to the use of the underlying 11.3/12.7 and 11.3/6.2~$\mu$m continuum ratios: (cont(11.3~$\mu$m)/cont(6.2~$\mu$m)$>$ 2.5 and cont(11.3~$\mu$m)/cont(12.7~$\mu$m)$>$ 2.5). However, we note they these are affected by the extinction coming from the host galaxy as well as the nuclear region, whereas the foreground extinction is cancelled out in the PAH EW ratios. Therefore, we suggest that the selection based on continuum ratios should only be used when the 6.2, 11.3 and/or 12.7~$\mu$m PAH bands are not detected.\\

\item  By using the PAH equivalent width ratio diagram, we classify as CON candidates 14 ULIRGs and 10 LIRGs from local U/LIRGs samples such as the HERUS and GOALS samples. We find that our diagnostic method identifies 30\% ULIRGs and 7\% LIRGs as CON candidates. Using the lower detection rate, we infer a CON density $\gtrsim$4.1$\times$10$^{-6}$ Mpc$^{-3}$.\\ 

\item  The observed continuum ratios of CON-dominated sources can be matched assuming torus models with a tapered disk geometry and a smooth dust distribution. This suggests that the nuclear dusty structure of deeply obscured galaxy nuclei has an extremely high dust coverage.\\

\item We find that the use of mid-IR color–color diagrams is an effective method to select CON-dominated sources at different redshifts. In particular, the combination of filters of the JWST/MIRI will enable the selection of CONs out to z$\sim$1.5. This will allow extending the selection of CONs to high redshifts where U/LIRGs are more numerous.\\

The high spatial resolution (improvement of a factor of 10 in spatial resolution with respect to Spitzer/IRS) and unprecedented sensitivity that will be afforded by the JWST will allow observations of distant and faint galaxies containing CONs. In the future, Origins Space Telescope-like missions will provide the high sensitivity and IR spectral coverage needed to extrapolate our mid-IR colour-colour technique to identify CONS amongst more distant galaxy populations.

   \end{enumerate}

\begin{acknowledgements}
IGB and DR acknowledge support from STFC through grant ST/S000488/1. DR also acknowledges support from the University of Oxford John Fell Fund.
The authors thank Almudena Alonso-Herrero and Miguel Pereira-Santaella for discussions, and Jason Marshall for providing the galaxy disk template spectrum. This work is based [in part] on archival data obtained with the Spitzer Space Telescope, which is operated by the Jet Propulsion Laboratory, California Institute of Technology under a contract with NASA. This research has also made use of the NASA/IPAC Extragalactic Database (NED), which is operated by the Jet Propulsion Laboratory, California Institute of Technology under a contract with NASA.

Finally, we thank the anonymous referee for their useful comments.

\end{acknowledgements}


\begin{appendix}
\section{Fraction of the galaxy covered by Spitzer/IRS}
\label{fraction_irs}
Here we investigate the fraction of the galaxy integrated mid-IR emission that is measured by Spitzer/IRS. We remark that the area of the galaxy covered by the slit of Spizer/IRS depends on the particular size of the galaxy and its distance. To do so, we retrieve 12~$\mu$m IRAS fluxes from the Infrared Astronomical Satellite (IRAS) Revised Bright Galaxy Sample (RBGS; \citealt{Sanders03}) and the IRAS Point Source Catalog{\footnote{https://irsa.ipac.caltech.edu/IRASdocs/surveys/psc.html}} (IRAS PSC; \citealt{Helou86}). Note that IRAS data were obtained with a 0.6-m telescope, with a pixel size of 2\arcmin ~and an angular resolution of $\sim$4\arcmin~. Therefore, the fluxes measured within the large IRAS apertures are representative of the integrated properties of the galaxies. Using the IRAS and Spitzer/IRS flux ratio at 12~$\mu$m, we can estimate the fraction of the total galaxy mid-IR emission measured by Spitzer/IRS.

\begin{figure}
\centering
\par{
\includegraphics[width=8.72cm]{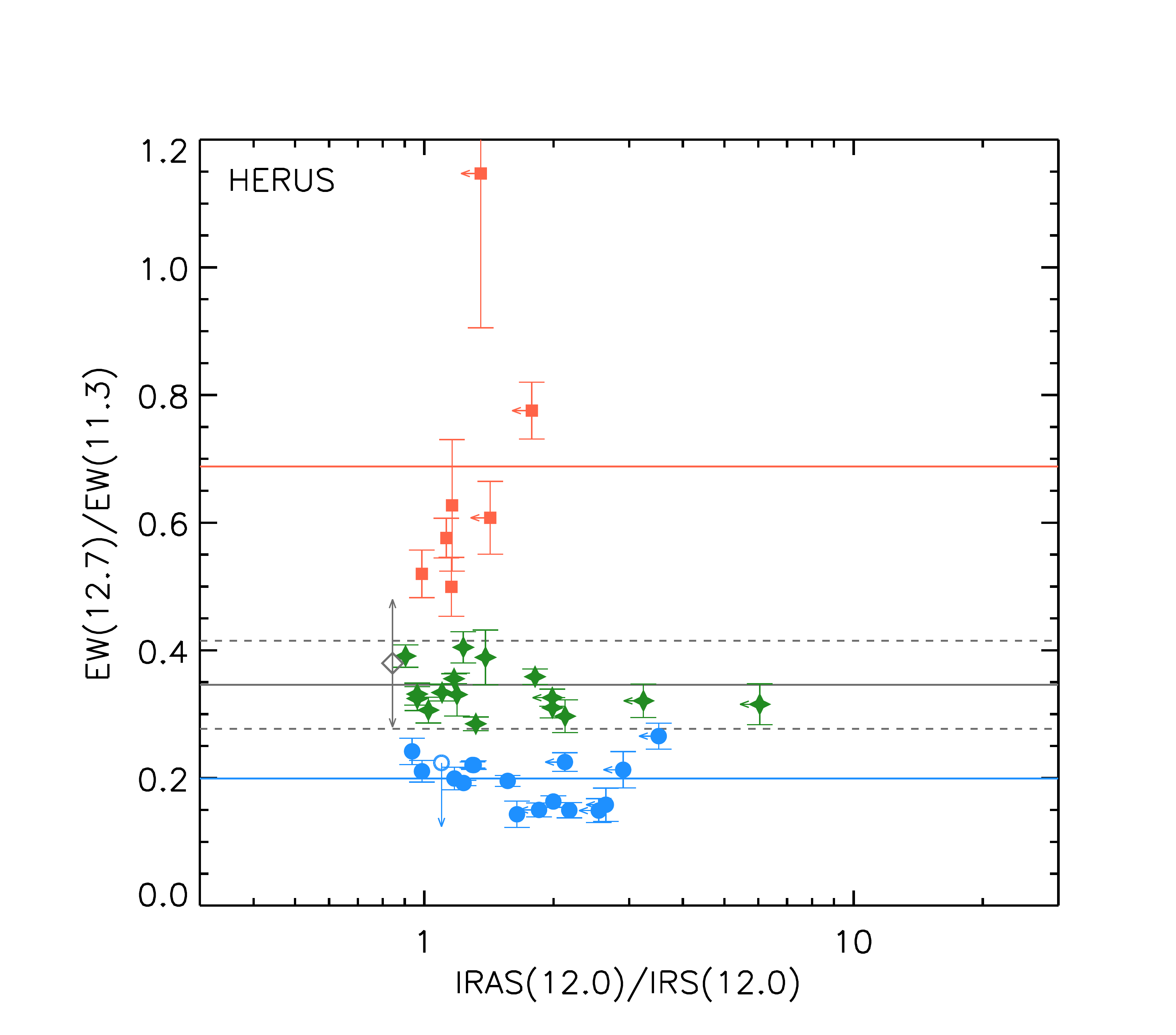}
\includegraphics[width=8.72cm]{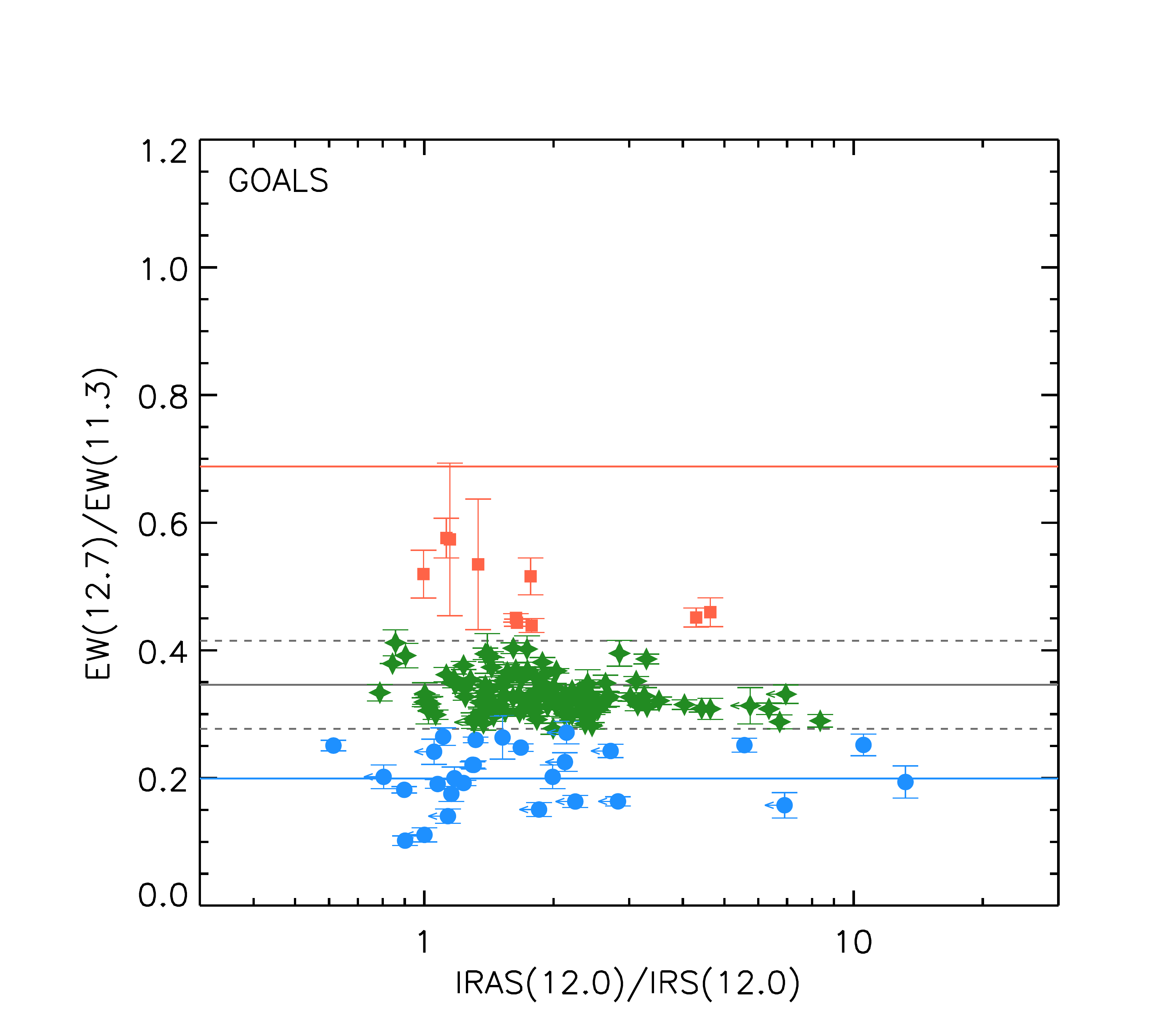}
\par}
\caption{Top panel: 12.7/11.3~$\mu$m PAH EW ratio versus the IRAS 12~$\mu$m/IRS 12~$\mu$m flux ratios for the HERUS sample. Green, blue and red symbols correspond to groups 1, 2 and 3 of sources (see Section \ref{127}. Bottom panel: same as the top panel but using the GOALS sample. The grey solid and dashed horizontal lines correspond to the average and 3$\sigma$ values found by \citet{Hernan-Caballero20} for star-forming galaxies. The blue and red solid horizontal lines represent the average values found for groups 2 and 3 of ULIRGs (see Section \ref{127}).}
\label{fractions}
\end{figure}

Fig. A.1 shows that the classification of groups 1 and 2 of ULIRGs does not depend on the fraction of the total galaxy measured by Spitzer/IRS. We find that ULIRGs tend to have smaller IRAS 12~$\mu$m/IRS 12~$\mu$m flux ratios (average value of 1.32$\pm$0.39 and 1.13$\pm$0.27 for the HERUS and GOALS ULIRGs, respectively) than LIRGs (average value of 5.11$\pm$12.79 for the GOAL LIRGs). This is in agreement with previous works (e.g. \citealt{Diaz-Santos10}) which showed that the mid-IR continuum emitting regions of ULIRGs are compact and less than 30\% of their mid-IR emission is extended. However, in general, LIRGs are not as compact as ULIRGs (e.g. \citealt{Diaz-Santos08}, \citealt{Diaz-Santos10}).

\end{appendix}

\end{document}